\documentclass[10pt, doublecolumn]{IEEEtran}
\usepackage{epsfig,latexsym}
\usepackage{float}
\usepackage{indentfirst}
\usepackage{amsmath}
\usepackage{bm}
\usepackage{amssymb}
\usepackage{times}
\usepackage{enumitem}
\usepackage{cite}
\usepackage{lastpage}
\usepackage{color} 
\usepackage{amsthm}
\usepackage{array}
\usepackage{algorithm}
\usepackage{algorithmic}
\usepackage{multirow}

\begin{document}
\title{Secure Multicast Communications with Pinching-Antenna Systems (PASS)}
\author{Shan Shan, Chongjun Ouyang, Yong Li, and Yuanwei Liu
\thanks{Shan Shan and Yong Li are with the School of Information and Communication Engineering, Beijing University of Posts and Telecommunications, Beijing 100876, China (email: \{shan.shan, liyong\}@bupt.edu.cn). Chongjun Ouyang is with the School of Electronic Engineering and Computer Science, Queen Mary University of London, London E1 4NS, U.K. (email: c.ouyang@qmul.ac.uk). Yuanwei Liu is with the Department of Electrical and Electronic Engineering, The University of Hong Kong, Hong Kong (email: yuanwei@hku.hk).}
}
\IEEEaftertitletext{\vspace{-3em}}
 \maketitle
\begin{abstract} 
This article investigates secure multicast communications in pinching-antenna systems (PASS), where \emph{pinching beamforming} is enabled by adaptively adjusting pinching antenna (PAs) positions along waveguides to improve multicast security.
Specifically, a PASS-based secure multicast framework is proposed, in which joint optimization of transmit and pinching beamforming is conducted to maximize the secrecy multicast rate.
i) For the single-group multicast scenario, an alternating optimization (AO) framework is employed, where the pinching beamformer is updated via an element-wise sequential optimization method. The transmit beamformer is designed via a semidefinite relaxation (SDR) formulation for an upper-bound solution, while a Dinkelbach-alternating direction method of multipliers (ADMM) offers a low-complexity alternative.
ii) For the multi-group multicast scenario, transmit and pinching beamformers are alternately optimized under a majorization-minimization (MM) framework. The transmit beamformer is obtained via SDR or an efficient second-order cone programming (SOCP) method, while the pinching beamformer is updated through MM-based element-wise sequential update strategy.
Numerical results are provided to demonstrate that: (i) PASS consistently outperform conventional fixed-location antenna architectures in terms of secrecy performance across various configurations; and (ii)  the performance advantage of PASS over fixed-location architectures becomes more significant with increased service region, larger antenna arrays, and higher user and eavesdropper densities. 
\end{abstract}


\section{Introduction}
Wireless communication systems have witnessed rapid advancements in recent years, driven by the growing demand for higher data rates, robust connectivity, and secure transmission. Multi-antenna technologies have played a critical role in achieving these goals by introducing spatial degrees of freedom (DoFs), which facilitate beamforming and spatial multiplexing to enhance both reliability and spectral efficiency~\cite{1386525}.
However, conventional multi-antenna systems typically adopt fixed-location antenna configurations. This rigid physical structure limits the ability to adapt to dynamic wireless environments, such as user mobility, signal blockage, or time-varying interference. To overcome this limitation, the concept of flexible-antenna systems has emerged, which enables wireless channel reconfigurability in real time. Representative approaches include reconfigurable intelligent surfaces (RISs)\cite{8741198}, movable antennas~\cite{10286328}, and fluid antennas~\cite{9264694}. RISs manipulate signal reflection phases to improve propagation conditions. Movable antennas change their physical positions to modify link geometry, while fluid antennas reshape their electromagnetic aperture through conductive fluid redistribution. These architectures offer potential benefits in mitigating small-scale fading by adjusting the propagation path locally. However, their effectiveness is generally constrained by limited movement ranges, which restricts their capability in addressing large-scale path loss or line-of-sight (LoS) blockage.

Recently, \emph{Pinching-Antenna Systems} (PASS) have been validated as the deployable flexible-antenna architecture to overcome the aforementioned limitations~\cite{yuanwei2025overview, yang2025pinching}. Originally introduced by NTT DOCOMO~\cite{fukuda2022pinching}, PASS utilize low-propagation-loss dielectric waveguides to deliver  electromagnetic signals, along which discrete dielectric particles--referred to as \emph{pinching antennas (PAs)}--are attached. These PAs act as reconfigurable radiation elements that can be mechanically or electrically activated to emit signals from arbitrary positions along the waveguide into free space. By adjusting PA positions based on user distribution, PASS enable reconfiguration of the spatial propagation environment, which is a capability we refer to as \emph{pinching beamforming}. Conceptually, PASS are also compatible with the emerging paradigm of surface-wave communication superhighways~\cite{10643519, 10742352}, which utilize reconfigurable waveguides to support in-waveguide signal propagation, reduce path loss, and improve delivery efficiency.
 This unique architecture provides several advantages over existing flexible-antenna solutions. 
 1) First, by enabling radiation at arbitrarily reconfigurable positions along an extensive dielectric waveguide, PASS can establish stable LoS links even in obstructed environments. Since most of the signal propagates within the low-loss dielectric medium, the architecture minimizes energy leakage and improves robustness against LoS blockage~\cite{11036558}. 
 2) Second, the flexible installation and repositioning of PAs allows for scalable deployment. With the specific pinching beamforming strategy, PASS enable dynamic reconfiguration of the antenna aperture geometry as well as wireless channel.
 3) Third, the enhanced spatial selectivity enabled by PASS allows precise energy focusing toward intended users while suppressing unintended signal leakage, which facilitates more effective interference management and strengthens the secure transmission.

\subsection{Prior Works}
Given these promising advantages, PASS have attracted increasing research interest in terms of performance characterization. Early studies have established foundational frameworks to understand its signal behavior and radiation mechanism. For instance, The pioneering theoretical study in~\cite{10945421} analyzed the fundamental transmission characteristics of PASS, followed by~\cite{10981775} which analyzed the array gain of multi-PA systems and derived optimal PA deployment strategies in terms of number and PA spacing. The influence of LoS blockage was revealed in~\cite{11036558}. The studies in~\cite{10976621} incorporated waveguide attenuation to evaluate outage probability and average rate under lossy propagation. From a signal modeling perspective, the authors in~\cite{wang2025modeling} introduced a physics-based directional coupler model for PA design, which derived two wave-coupled power models based on electromagnetic theory. An electromagnetic-compliant channel model for orthogonal frequency division multiplexing (OFDM)-based PASS was introduced in\cite{11114424} to characterize frequency-selective and dispersion effects near the waveguide cutoff. The channel estimation for PASS was firstly explored in~\cite{11018390}, which addressed their uniquely ill-conditioned and underdetermined channel characteristics.

Based on these fundamental performance analysis, optimizing PA positions also plays a critical role in PASS. Specifically, the authors in\cite{11050939} introduced a probability-learning algorithm for PA placement, and \cite{10896748} employed a two-stage position adjustment strategy for downlink PASS. The authors in~\cite{11063467} explored graph neural network (GNN)-enabled joint optimization for PA placement and power allocation, while the authors in~\cite{11048566} designed two-timescale beamforming strategies. In non-orthogonal multiple access (NOMA)-assisted scenarios, a matching-based optimization was proposed in\cite{10912473}, while closed-form placement solutions for both orthogonal multiple access (OMA) and NOMA schemes was derived in~\cite{11016750}. The work in~\cite{10909665} further decoupled PA positioning from resource allocation to optimize the minimum uplink rate, and the authors in~\cite{11029492} investigated NOMA-assisted downlink PASS.
Moreover, PASS have been applied to a wide range of wireless communication scenarios, such as wireless sensing~\cite{11111701}, simultaneous wireless information and power transfer (SWIPT)~\cite{11106459}, and Internet-of-Things (IoT)-based MIMO systems~\cite{11098708}. Other studies addressed challenges in wireless powered networks~\cite{11096622}.

Owing to the capability of adjusting PAs along arbitrarily deployed waveguides, PASS can exploit spatial selectivity to enlarge the channel disparity between intended and unintended receivers, which offers inherent advantages for enhancing physical-layer security (PLS) over conventional fixed-location antenna arrays.  In particular, joint optimization of transmit and pinching beamforming was developed in~\cite{Osamah2025pls, Kaidi2025pls} to maximize secrecy rates, while PA activation and baseband beamforming algorithms for both single-user and multi-user settings was designed in~\cite{Mingjun2025pls}. Artificial-noise (AN)-aided designs were investigated in~\cite{ Pigi2025pls}, and the authors in~\cite{Hao2025covert} proposed a PASS-assisted covert communication framework to minimize detection probability by adaptively shaping the radiation footprint.

While most existing PASS-based secrecy studies have concentrated on unicast transmission, secure multicast communication is becoming increasingly important in future wireless networks to support group-oriented applications such as content broadcasting and federated learning. Unlike unicast, multicast secrecy requires the confidential delivery of a common message to all intended users, which imposes both fairness and secrecy constraints. In this setting, system performance is inherently limited by the user with the weakest channel conditions. Moreover, the presence of multiple eavesdroppers further exacerbates the problem, which makes the resulting optimization highly non-convex. PASS offer a promising solution by exploiting its spatial selectivity--enabled through arbitrarily deployed waveguides and flexible PA activation--to enlarge the channel disparity between intended and unintended receivers, which strengthens multicast secrecy performance beyond what is achievable with conventional fixed-location antenna arrays.

\subsection{Motivations and Contributions}
Despite this advantage, the potential of PASS for secure multicast transmission remains largely under explored. To address this gap, this paper investigates the secrecy application of PASS in the multicast transmission scenario. 
The key contributions of this work are summarized as follows:
\begin{itemize}
    \item We propose a secure transmission framework for multicast communications in PASS, applicable to both single- and multi-group scenarios with one or more groups of legitimate users and a common set of eavesdroppers. To ensure rate fairness and robust group performance, we formulate a secrecy multicast rate maximization problem by jointly optimizing transmit and pinching beamforming.
    
    \item For the single-group secure multicast scenario, we develop an alternating optimization (AO) algorithm that iteratively updates the transmit and pinching beamformers. We update the pinching beamformer using a sequential element-wise method. To handle the non-convex fractional objective in the transmit beamforming subproblem, we first apply the Charnes-Cooper transformation and then perform semidefinite relaxation (SDR) to obtain an upper-bound solution. To reduce complexity, we further develop a low-complexity method based on the Dinkelbach framework, where each subproblem is efficiently solved via alternating direction method of multipliers (ADMM) approach.

    \item For the multi-group secure multicast scenario, we adopt a majorization-minimization (MM) framework to address the non-convexity. Within this framework, the pinching beamformer is optimized using an MM-based sequential element-wise method, while the transmit beamformer subproblem is solved via SDR to obtain an upper-bound solution. To further reduce complexity, we develop a low-complexity alternative based on second-order cone programming (SOCP).

    \item Extensive simulation results validate the proposed framework. PASS consistently outperform conventional fixed-location antenna systems in terms of multicast secrecy performance, with the performance gain becoming more significant in larger coverage areas and under higher densities of legitimate and eavesdropping users. Furthermore, the proposed low-complexity Dinkelbach-ADMM method for the single-group case and the SOCP solution for the multi-group case achieve secrecy multicast rates that are comparable to those obtained by the SDR approach. This confirms the effectiveness of these algorithms in achieving a favorable balance between computational complexity and secrecy performance.
\end{itemize}

\subsection{Organization and Notations}
The remainder of this paper is organized as follows. Section~\ref{System_Model} introduces the PASS-based secure multicast transmission system and formulates the secrecy multicast rate maximization problem. 
Sections~\ref{Single_group} and \ref{Multiple_group} present the joint multicast secure beamforming design for the single- and multiple-group scenarios, respectively. Numerical results are presented in Section~\ref{simulation}. Finally, Section~\ref{conclusion} concludes this paper.

\subsubsection*{Notations}
Scalars, vectors, and matrices are represented by regular, bold lowercase, and  bold uppercase letters, respectively. The sets of complex numbers, real numbers and Hermitian matrices are denoted by $\mathbb{C}$, $\mathbb{R}$ and ${\mathbb H}$. The inverse, conjugate, transpose, conjugate transpose, and trace operators are denoted by $(\cdot)^{-1}$, $(\cdot)^{\ast}$, $(\cdot)^{\rm T}$, $(\cdot)^{\rm H}$, and ${\rm Tr}(\cdot)$, respectively. The rank of a matrix is denoted by $\mathrm{Rank}(\cdot)$. For a vector ${\bf x}$, $[{\bf x}]_i$ denotes its $i$th element. ${\rm{Diag}}({\bf x})$ denotes the diagonal matrix with the diagonal elements formed by the vector ${\bf x}$. ${\bf X} \succeq 0$ denotes that matrix ${\bf X}$ is positive semidefinite. ${\rm{BlkDiag}}([{{\bf A}};{{\bf B}}])$ denotes the block-diagonal matrix formed by the matrices ${\bf A}$ and ${\bf B}$. The operator $[x]^+ \triangleq \max\{x, 0\}$ denotes the non-negative projection of $x$. ${\mathcal C}{\mathcal N}(a, b^2)$ denotes a circularly symmetric complex Gaussian distribution with mean $a$ and variance $b^2$. The statistical expectation operator is represented by ${\mathbb{E}}\{\cdot\}$. The absolute value and Euclidean norm are denoted by $|\cdot|$ and $\|\cdot\|$, respectively. The real part of a complex number is denoted by $\Re \{\cdot\}$. The big-O notation is denoted by ${\mathcal O}(\cdot)$. 

\section{System Model and Problem Formulation} \label{System_Model}
\begin{figure}[!t]
\centering
\includegraphics[height=0.25\textwidth]{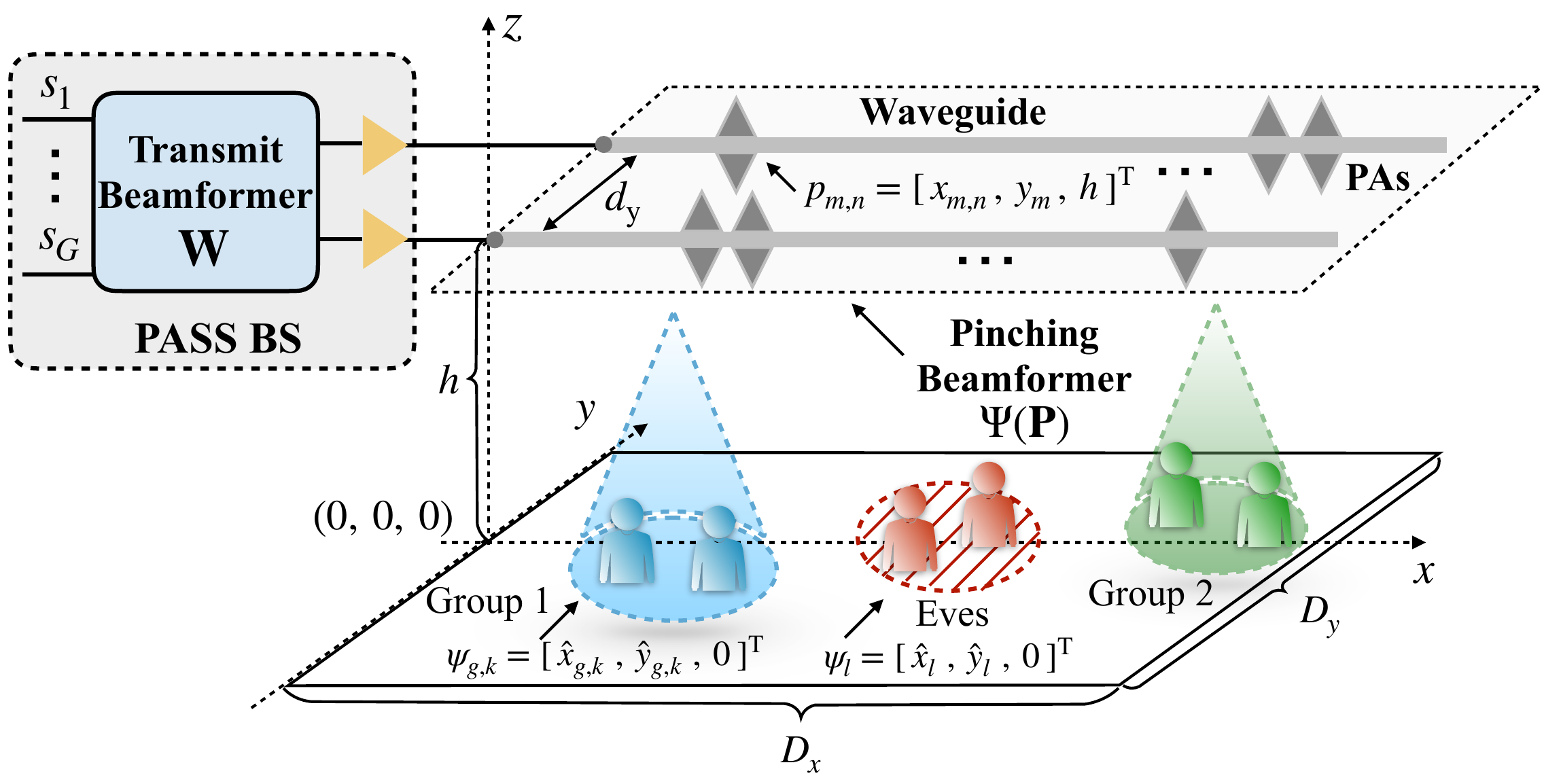}
\caption{Illustration of PASS-based secure multicast transmission architecture.}
\label{Fig_1}
\vspace{-10pt}
\end{figure}
We consider downlink secure multicast transmission enabled by PASS, as illustrated in {\figurename}~{\ref{Fig_1}}. The base station (BS) is equipped with $M$ parallel waveguides, each activated by $N$ PAs, to serve $G$ multicast groups, where $M\geq G\geq 1$. Both the waveguides and the PAs are installed at a fixed height of $h$ within a rectangular coverage area of size $D_{\rm x} \times D_{\rm y}$. The waveguides are uniformly spaced along the $y$-axis with an interval of $d_{\rm y} = D_{\rm y}/(M-1)$. 
The feed point of the $m$th waveguide is located at ${p}_{m,0} = \left[0, y_{m},  h\right]^{\rm T}$, and the position of its $n$th PA is given by ${p}_{m,n} = \left[x_{m,n},  y_{m},  h\right]^{\rm T}$, where $x_{m,n} \in \mathcal{S}_x$ denotes the candidate PA locations along the $x$-axis. 
Let ${\mathcal M}\triangleq\{1,\dots,M\}$ and ${\mathcal N}\triangleq\{1,\dots,N\}$ be the index sets of all waveguides and PAs per waveguide, respectively. The $x$-coordinates of the PAs on waveguide $m$ are collected in the vector ${\bf p}_{m} \triangleq [x_{m,1}, x_{m,2}, \dots, x_{m,N}]^{\rm T}$, which satisfy 
$0 \le x_{m,1} < \dots < x_{m,N} \le D_{{\rm x}}$ and a minimum inter-PA spacing constraint $|x_{m,n} - x_{m,n-1}| \ge \Delta_{\min} = \lambda / 2$ for $n \ge 2$, where $\lambda$ is the free-space wavelength~\cite{10981775}. 
The complete PA location matrix is denoted by ${\bf P} \triangleq [{\bf p}_{1}, \dots, {\bf p}_{M}] \in {\mathbb R}^{M \times N}$.

In the considered secure multicast system, the legitimate users (Bobs) are partitioned into $G$ disjoint multicast groups, $\mathcal{K}_g$, $g \in \mathcal{G} \triangleq \{1,\dots,G\}$, such that $\bigcup_{g=1}^G \mathcal{K}_g = \{1,\dots,K\}$ and $\mathcal{K}_i \cap \mathcal{K}_j = \emptyset$ for $i \ne j$. 
All users in the $g$th group receive a common confidential message. 
The location of the $k$th user in group $g$ is $\hat{\boldsymbol\psi}_{g,k} = [\hat{x}_{g,k}, \hat{y}_{g,k}, 0]^{\mathrm{T}}$. 
Moreover, the system is subject to interception from $L$ external eavesdroppers (Eves), whose positions are denoted by $\hat{\boldsymbol\psi}_{\mathrm{e},l} = [\hat{x}_{{\rm e},l}, \hat{y}_{{\rm e},l}, 0]^{\rm T}$, $l \in \mathcal{L} \triangleq \{1, \dots, L\}$.
\vspace{-5pt}
\subsection{System Model}
The original group-based confidential signals are first processed at baseband via the transmit beamforming matrix, upconverted by the radio frequency (RF) chains, and then fed into the waveguides for in-waveguide propagation before being radiated into free space. The transmitted signal radiated by all PAs can be written as follows:
\begin{align}
  {\bf x} = \mathbf{\Psi}({\bf P}) \sum_{g=1}^{G} {\bf w}_g s_g,
\end{align}
where $s_g \in \mathbb{C}$ represents the Gaussian data symbol for group $g$ with $\mathbb{E}[|s_g|^2]=1$, and ${\bf w}_g \in {\mathbb C}^{M\times 1}$ denotes the transmit beamforming vector for the $g$th group. The pinching beamforming matrix $\mathbf{\Psi}({\bf P}) \in {\mathbb C}^{MN \times M}$ characterizes the in-waveguide signal propagation from waveguide feed points to their associated active PA, 
which can be modeled as a block diagonal matrix
\begin{align}
  \mathbf{\Psi}({\bf P}) 
  = \text{BlkDiag} \big( [\boldsymbol{\psi}({\bf p}_1); \dots ; \boldsymbol{\psi}({\bf p}_M)] \big).
\end{align}
Here, $\boldsymbol{\psi}({\bf p}_m) \in \mathbb{C}^{N \times 1}$ denotes the in-waveguide propagation vector from the $m$th waveguide to its $N$ associated PAs. The $n$th element of $\boldsymbol{\psi}({\bf p}_m)$ is given by
\begin{align}
  \left[ \boldsymbol{\psi}({\bf p}_m) \right]_n 
  = \sqrt{P_{m,n}} \exp \left( -{\rm j} k_{\rm g} x_{m,n} \right),
\end{align}
where $P_{m,n}$ denotes the $(m,n)$th PA power coefficient. In this paper, we assume equal power allocation across the $N$ PAs to simplify modeling and optimization, which implies $P_{m,n} = \frac{1}{N}$~\cite{wang2025modeling}. Moreover, $k_{\rm g} = 2\pi / \lambda_{\rm g}$ is the guided wavenumber and $\lambda_{\rm g} = \lambda / n_{\text{eff}}$ is the guided wavelength with $n_{\text{eff}}$ being the effective refractive index of the dielectric waveguide~\cite{10945421}.

Since PASS are highly compatible with high-frequency~\cite{fukuda2022pinching}, we adopt a free-space LoS-based channel model for both Bobs and Eves. The channel from PASS to the $(g,k)$th Bob is
\begin{align}
  {\bf h}_{g,k}({\bf P}) 
  &= \left[ {\bf h}^{\rm T}_{g,k}({\bf p}_1), \ldots, {\bf h}^{\rm T}_{g,k}({\bf p}_M) \right]^{\rm T},
\end{align}
where ${\bf h}_{g,k}({\bf p}_m) \in \mathbb{C}^{N \times 1}$ has entries
\begin{align}
  \left[{\bf h}_{g,k}({\bf p}_m)\right]_n 
  = \frac{\sqrt{\eta} \exp\left(-{\rm j}k_0\|\bm\psi_{m,n}-\hat{\bm\psi}_{g,k}\|\right)}
          {\|\bm\psi_{m,n}-\hat{\bm\psi}_{g,k}\|},
\end{align}
with $k_0 = 2\pi/\lambda$ denoting the free-space wavenumber, and $\eta = c^2/(16\pi^2 f_{\rm c}^2)$ collecting system constants including the carrier frequency $f_{\rm c}$ and the speed of light $c$~\cite{10945421}.

Similarly, the channel response of the $l$th Eve is given by
\begin{align}
  {\bf h}_{{\rm e},l}({\bf P}) 
  &= \left[ {\bf h}^{\rm T}_{{\rm e}, l}({\bf p}_1), \ldots, {\bf h}^{\rm T}_{{\rm e}, l}({\bf p}_M) \right]^{\rm T},
\end{align}
where
\begin{align}
  \left[{\bf h}_{{\rm e}, l}({\bf p}_m)\right]_n 
  = \frac{\sqrt{\eta} \exp\left(-{\rm j}k_0\|\bm\psi_{m,n}-\hat{\bm\psi}_{{\rm e}, l}\|\right)}
          {\|\bm\psi_{m,n}-\hat{\bm\psi}_{{\rm e}, l}\|}.
\end{align}
Both the in-waveguide matrix $\mathbf{\Psi}({\bf P})$ and the radiation channels ${\bf h}_{g,k}({\bf P})$, ${\bf h}_{{\rm e}, l}({\bf P})$ depend solely on ${\bf P}$, which enables a compact effective channel representation as follows:
\begin{align}
  \hat{{\bf h}}_{g,k}^{\rm T}({\bf P}) 
  &= {\bf h}_{g,k}^{\rm T}({\bf P})\,\mathbf{\Psi}({\bf P}), \\
  \hat{{\bf h}}_{{\rm e}, l}^{\rm T}({\bf P}) 
  &= {\bf h}_{{\rm e}, l}^{\rm T}({\bf P})\,\mathbf{\Psi}({\bf P}).
\end{align}

The received signal at the $(g,k)$th Bob is
\begin{align}
  y_{g,k} 
  = \hat{\bf h}_{g,k}^{\rm T}({\bf P}) \sum_{i=1}^{G} {\bf w}_{i} s_{i} + z_{g,k},
\end{align}
where $z_{g,k} \sim \mathcal{CN}(0, \sigma_{g,k}^2)$ is the additive Gaussian noise. 
The corresponding signal-to-interference-plus-noise ratio (SINR) is
\begin{align}\label{sinr_bob}
  {\rm SINR}_{g,k} 
  = \frac{|\hat{{\bf h}}_{g,k}^{\rm T}({\bf P}) {\bf w}_{g}|^2}
         {\sum_{i \ne g} |\hat{{\bf h}}_{g,k}^{\rm T}({\bf P}) {\bf w}_{i}|^2 + \sigma_{g,k}^2}.
\end{align}

For each Eve $l \in \mathcal{L}$, the SINR when attempting to decode the $g$th group message can be written as follows:
\begin{align}\label{sinr_eve}
  {\rm SINR}_{g,l} 
  = \frac{|\hat{{\bf h}}_{{\rm e},l}^{\rm T}({\bf P}) {\bf w}_g|^2}
         {\sum_{i \ne g} |\hat{{\bf h}}_{{\rm e},l}^{\rm T}({\bf P}) {\bf w}_i|^2 + \sigma_{{\rm e},l}^2}.
\end{align}

Under the worst-case Eve assumption, the per-user secrecy rate can be defined as follows:
\begin{align}\label{eq:per-user-secrecy-rate}
  R_{g,k}^{\sec} 
  = \left[ \log_2(1 + {\rm SINR}_{g,k}) - \max_{l \in \mathcal{L}} \log_2(1 + {\rm SINR}_{g,l}) \right]^+,
\end{align}
and the secrecy multicast rate for the $g$th group is
\begin{align}
  R^{\sec}_g(\{{\bf w}_i\}_{i = 1}^{G},\, {\bf P}) 
  = \min_{k \in \mathcal{K}_g} R_{g,k}^{\sec}.
\end{align}

\subsection{Problem Formulation}
We aim to maximize the secrecy multicast rate of the considered PASS-aided system by jointly optimizing the transmit beamforming vectors $\{{\bf w}_i\}_{i=1}^{G}$ and the PA location matrix ${\bf P}$. The optimization problem is formulated as follows:
\begin{subequations}\label{eq:Problem-secure}
\begin{align}
  &\max_{\{{\bf w}_i\}_{i=1}^{G},\,{\bf P}} 
    \ \min_{g \in \mathcal{G}} \ R^{\sec}_g\big(\{{\bf w}_i\}_{i=1}^{G},\,{\bf P}\big) \\
  &\text{s.t.} \quad
    \sum_{i=1}^{G} \|{\bf w}_i\|^2 \le P_{\rm t}, \label{power_constraint} \\
  &\quad \ \ \ x_{m,n} \in \mathcal{S}_x, \quad \forall m \in \mathcal{M}, \ n \in \mathcal{N}, \label{S_x} \\
  &\quad \ \ \ |x_{m,n+1} - x_{m,n}| \ge \Delta_{\min}, \quad \forall m \in \mathcal{M}, \ n \ge 1, \label{delta_min}
\end{align}
\end{subequations}
where $P_{\rm t}$ in~\eqref{power_constraint} denotes the maximum transmit power, and constraints~\eqref{S_x}-\eqref{delta_min} specify the feasible PA placement set and the minimum inter-PA spacing requirement, respectively.

Problem~\eqref{eq:Problem-secure} is highly challenging due to its \emph{non-convex}, \emph{non-smooth}, and \emph{strongly coupled} structure. 
First, the objective involves a nested max-min secrecy rate expression across all Bob-Eve pairs, which results in a non-smooth function that is difficult to optimize directly. 
Second, the beamforming vectors $\{{\bf w}_i\}$ and the PA positions ${\bf P}$ are multiplicatively coupled in both Bob and Eve channel responses, which leads to a highly non-convex feasible set. 
Since there are no standard methods for obtaining a globally optimal solution to such problems, we adopt an AO framework that decomposes the joint problem into more tractable subproblems and solves them iteratively.

\section{Joint Secure Beamforming for Single-Group Multicast Communications}
\label{Single_group}

In this section, we consider the single-group secure multicast problem, where $M$ waveguides jointly transmit a common confidential message $s \sim \mathcal{CN}(0,1)$ to $K$ Bobs in the presence of $L$ Eves.  
Let ${\bf w} \in \mathbb{C}^{M\times 1}$ denote the transmit beamforming vector, subject to the total power constraint $\|{\bf w}\|^2 \le P_{\rm t}$. The received signals at the $k$th Bob and the $\ell$th Eve are respectively given by
\begin{align}
	y_{b,k}={\bf\hat h}_{k}^{\rm T}({\bf P}){\bf w}\,s + z_{k},\quad
y_{{\rm e},\ell}={\bf\hat h}_{{\rm e},\ell}^{\rm T}({\bf P}) {\bf w}\,s + z_{{\rm e},\ell}.
\end{align}
where $\hat{{\bf h}}_{k}({\bf P})$ and $\hat{{\bf h}}_{{\rm e},\ell}({\bf P})$ denote the effective PASS-to-Bob and PASS-to-Eve channel vectors, respectively, and $z_{k} \sim \mathcal{CN}(0,\sigma_{k}^2)$, $z_{{\rm e},\ell} \sim \mathcal{CN}(0,\sigma_{{\rm e},\ell}^2)$ represent the additive Gaussian noise.
According to~\eqref{eq:per-user-secrecy-rate}, the achievable secrecy multicast rate for the single-group scenario can be written as follows:
\begin{align}\label{eq:Cs_def}
R^{\sec}({\bf P}, {\bf w})
&= \left[ 
\min_{k \in \{1,\dots,K\}} \log_2\left( 1 + \frac{|\hat{{\bf h}}_{k}^{\rm T}({\bf P}){\bf w}|^2}{\sigma_k^2} \right) \right. \notag \\
& - \left.
\max_{\ell \in \{1,\dots,L\}} \log_2\left( 1 + \frac{|\hat{{\bf h}}_{\ell}^{\rm T}({\bf P}){\bf w}|^2}{\sigma_{{\rm e},\ell}^2} \right) 
\right]^+.
\end{align}

\subsection{Secure Transmit Beamforming Optimization}
In this subsection, we focus on optimizing the transmit beamformer ${\bf w}$ with the fixed pinching beamformer $\mathbf{\Psi}({\bf P})$. The original problem can be simplified as follows:
\begin{align}
  \max_{\bf w} \ R^{\sec}({\bf w}),
  \quad \text{s.t.} \ \|{\bf w}\|^{2} \le P_{\rm t}.
  \label{eq:TxBF_init}
\end{align}
However, the objective exhibits a nested max-min structure, which is both non-convex and non-smooth. 

\subsubsection{SDR Method} 
To tackle this challenge, we first adopt a SDR approach, which provides a tractable approximation and serves as a tight upper bound. Specifically, we drop the outer $\log(\cdot)$ operation and $[\cdot]^+$ operators to obtain a conservative reformulation, which facilitates fractional SDR modeling and convex optimization\footnote{The outer $\left[\cdot\right]^+$ operator is removed to obtain a conservative approximation, since maximizing a lower-bounded function naturally leads to solutions within the positive regime.}
:
\begin{align}
  \max_{\bf w} \frac{\min_{k} 1 + \sigma_{k}^{-2}|\hat{{\bf h}}_{k}^{\rm T}{\bf w}|^2}{\max_{\ell} 1 + \sigma_{{\rm e},l}^{-2}|\hat{{\bf h}}_{{\rm e},\ell}^{\rm T} {\bf w}|^2}, \quad {\rm s.t.}\ \|{\bf w}\|^2\leq P_{\rm t}.
  \label{eq:frac_ratio}
\end{align}
By changing the maximization to minimization and exchanging the numerator and denominator in~\eqref{eq:frac_ratio}, we obtain
\begin{align}\label{eq:sdr}
	\min_{\bf w} \ \frac{\max_{\ell} {\bf w}^{\rm H}\left(\frac{1}{M}{\bf I} + P_{{\rm eff}, \ell}{\bf H}_{\ell}\right){\bf w}}
	{\min_{k} {\bf w}^{\rm H}\left(\frac{1}{M}{\bf I} + P_{{\rm eff}, k}{\bf H}_{k}\right){\bf w}},
	\quad \text{s.t.} \ \|{\bf w}\|^2\leq 1,
\end{align}
where $P_{{\rm eff}, \ell}\triangleq P_{\rm t}/\sigma_{\ell}^2$, $P_{{\rm eff}, k}\triangleq P_{\rm t}/\sigma_{k}^2$, ${\bf H}_{\ell} \triangleq {\bf h}_{\ell}{\bf h}_{\ell}^{\rm H}$, and ${\bf H}_{k} \triangleq {\bf h}_{k}{\bf h}_{k}^{\rm H}$. The transmit beamformer has been normalized to unit norm for convenience.
Problem~\eqref{eq:sdr} can be addressed via SDR combined with the Charnes-Cooper transformation~\cite{5755208}. By introducing ${\bf W} = {\bf w}{\bf w}^{\rm H}$, problem~\eqref{eq:sdr} can be equivalently expressed as follows:
\begin{subequations}
\begin{align}\label{eq:convW}
	\min_{{\bf W}\in {\mathbb H}^{M}} & \ \frac{\max_{\ell}{\rm Tr}\left({\bf W}\left(\frac{1}{M}{\bf I} + P_{{\rm eff}, \ell}{\bf H}_{\ell}\right)\right)}
	{\min_k {\rm Tr}\left({\bf W}\left(\frac{1}{M}{\bf I} + P_{{\rm eff}, k}{\bf H}_{k}\right)\right)}, \\
	\text{s.t.} & \ {\rm Tr}({\bf W})\leq 1, \label{W_power_const}\\
	& \ {\bf W}  \succeq 0, \quad {\rm Rank}({\bf W}) = 1.
\end{align}
\end{subequations}

Dropping the rank-one constraint yields a fractional SDP, which can be converted into a standard SDP via the Charnes-Cooper transformation. By introducing $\widetilde{\bf W} = \zeta {\bf W}$ with $\zeta \geq 0$, the problem becomes
\begin{subequations}
	\begin{align}\label{eq:standardSDP}
		\min_{\widetilde{\bf W}\in {\mathbb H}^{M}, \zeta} \ & \max_{\ell} \ {\rm Tr}\left(\widetilde{\bf W}\left(\frac{1}{M}{\bf I} + P_{{\rm eff}, \ell}{\bf H}_{\ell}\right)\right), \\
		\text{s.t.} & \ \min_{k} \ {\rm Tr}\left(\widetilde{\bf W}\left(\frac{1}{M}{\bf I} + P_{{\rm eff}, k}{\bf H}_{k}\right)\right) = 1, \\
		& \ {\bf W}  \succeq 0, \ \eqref{W_power_const},
	\end{align}
\vspace{-5pt}
\end{subequations}
or equivalently,
\begin{subequations}\label{eq:standardSDP2}
	\begin{align}
		\min_{\widetilde{\bf W}\in {\mathbb H}^{M}, \gamma, \zeta} & \quad \gamma, \\
		\text{s.t.} & \ {\rm Tr}\left(\widetilde{\bf W}\left(\frac{1}{M}{\bf I} + P_{{\rm eff}, \ell}{\bf H}_{\ell}\right)\right) \leq \gamma, \ \forall \ell, \\
		& \ {\rm Tr}\left(\widetilde{\bf W}\left(\frac{1}{M}{\bf I} + P_{{\rm eff}, k}{\bf H}_{k}\right)\right) \geq 1, \ \forall k, \\
		& \ {\bf W}  \succeq 0, \ \zeta \geq 0, \ \eqref{W_power_const}.
	\end{align}
\end{subequations}

Problem~\eqref{eq:standardSDP2} is a standard convex SDR that can be efficiently solved using interior-point methods (e.g., CVX with SDPT3). If the resulting $\widetilde{\bf W}^{\star}$ is rank-one, the optimal beamformer is obtained as ${\bf w}^{\star} = \sqrt{\zeta^{-1}}\,{\bf w}_1$, where ${\bf w}_1$ is the principal eigenvector of $\widetilde{\bf W}^{\star}$. Otherwise, Gaussian randomization can be employed to construct a feasible beamforming vector~\cite{4567639}.

\subsubsection{Low-Complexity ADMM Method}
Although the SDR approach in~\eqref{eq:TxBF_init} yields tight performance for small $(K,L)$, its computational complexity $\mathcal{O}(M^{6})$ becomes prohibitive when $M$ or $(K,L)$ is large. To address this scalability issue, we develop a low-complexity non-convex optimization algorithm based on ADMM, which enables efficient beamformer updates without the semidefinite lifting required by SDR.
We first adopt a smooth log-sum-exp (LSE) approximation to handle the non-smooth max-min fractional structure in the objective function. Recall the identity~\cite{Boyd2004Convex}
\begin{align}
	  \max_i a_i = \lim_{\beta \to 0}\beta\,{\rm log}\sum_{i} e^{a_i/\beta},\quad \beta>0,
\end{align}
where $\beta>0$ is a smoothing factor. Applying it to \eqref{eq:Cs_def} yields a smooth fractional surrogate
\begin{align}\label{eq:smooth-obj}
  \min_{{\bf w}\in{\mathbb C}^M}
  \frac{f_1({\bf w})}{f_2({\bf w})} \quad\text{s.t. } \|{\bf w}\|^2\le P_{\rm t},
\end{align}
where\vspace{-5mm} 
\begin{align}
  f_1({\bf w}) &=\beta\log\sum_{\ell=1}^{L} \exp\left(\frac{1}{\beta}{\bf w}^{\rm H}\Big(\frac{1}{M}{\bf I}+P_{\rm t}{\bf h}_\ell{\bf h}_\ell^{\rm H}\Big){\bf w}\right),\\
  f_2({\bf w}) &= -\beta\log\sum_{k=1}^{K}\exp\left(-\frac{1}{\beta}{\bf w}^{\rm H}\Big(\frac{1}{M}{\bf I}+P_{\rm t}{\bf h}_k{\bf h}_k^{\rm H}\Big){\bf w}\right).
\end{align}

The objective in~\eqref{eq:smooth-obj} is a ratio of two quadratic functions, which falls within the scope of fractional programming. We adopt the classical Dinkelbach method to transform~\eqref{eq:smooth-obj} into a sequence of parameterized subproblems. At outer iteration $j$, we update
\begin{align}
  \varsigma^{(j)}  = \frac{f_1({\bf w}^{(j)})}{f_2({\bf w}^{(j)})},\quad
  {\bf w}^{(j+1)}  = \arg\min_{\|{\bf w}\|^2\le P_{\rm t}}
         \phi_{\varsigma^{(j)}}({\bf w}),\label{eq:eta_update}
\end{align}
with $\phi_{\varsigma}({\bf w})=f_1(\mathbf w)-\varsigma f_2({\bf w})$. Under standard regularity conditions, the sequence $\{\varsigma^{(j)}\}$ generated by~\eqref{eq:eta_update} is monotonically non-increasing and converges to the global optimum $\varsigma^\star$, while the optimality residual $\phi_{\varsigma^{(j)}}({\bf w}^{(j)}) \rightarrow 0$ as $j \rightarrow \infty$. The procedure is summarized in Algorithm~\ref{alg:dinkelbach}.

We next design an ADMM solver for the inner subproblem in~\eqref{eq:eta_update}. Introducing an auxiliary variable ${\bf u} \in \mathbb{C}^M$, the problem is equivalently written as follows:
\begin{algorithm}[t]
\caption{Dinkelbach Procedure for Problem~\eqref{eq:smooth-obj}}
\label{alg:dinkelbach}
\begin{algorithmic}[1]
\STATE Initialize a feasible ${\bf w}$ for~\eqref{eq:smooth-obj}
\REPEAT
    \STATE $\displaystyle 
       \varsigma \leftarrow
       \frac{f_1({\bf w})}{f_2({\bf w})}$
    \STATE $\displaystyle
       {\bf w} \;\leftarrow\;
       \arg\min_{\|{\bf w}\|^2\le P_{\rm t}}
       \bigl[f_1({\bf w})-\varsigma f_2({\bf w})\bigr]$
\UNTIL{stopping criterion is satisfied}
\STATE \textbf{output} ${\bf w}$
\end{algorithmic}
\end{algorithm}
\begin{subequations}\label{eq:phi_admm}
\begin{align}
  \min_{{\bf w}, {\bf u}\in\mathbb{C}^M} \  & \phi({\bf w}) \\
\text{s.t.} & \  \|{\bf w}\|^{2} \le P_{\rm t}, \ {\bf u} = {\bf w}.
\end{align}
\end{subequations}
This reformulation enables the use of ADMM to decompose the non-convex constrained problem into simpler subproblems that can be solved more efficiently in closed-form or with low complexity~\cite{8186925}. Let the augmented Lagrangian function be
\begin{align}
\mathcal{L}({\bf u},{\bf w},{\bm\nu}) 
= \phi({\bf w}) + \Re\{{\bm\nu}^{\rm H}({\bf u} - {\bf w})\} + \frac{\rho}{2}\|{\bf u} - {\bf w}\|^2.
\end{align}
To circumvent the difficulty caused by the non-quadratic $\phi({\bf w})$, we adopt a quadratic approximation for the ${\bf u}$- and ${\bf w}$-updates, which leads to the following iterative scheme in~\eqref{eq:ADMM_variant}, which is shown on the top of the next page. 
\begin{figure*}[!t]
\normalsize
\begin{subequations}\label{eq:ADMM_variant}
\begin{align}
{\bf u}^{(p+1)} &= \arg\min_{\|{\bf u}\|^{2} \le P_{\rm t}}\;
\mathcal{L}({\bf u}^{(p)},{\bf w}^{(p)},{\bm\nu}^{(p)}) 
+ \langle \nabla_{\bf u} \mathcal{L}({\bf u}^{(p)},{\bf w}^{(p)},{\bm\nu}^{(p)}), {\bf u} - {\bf u}^{(p)} \rangle
+ \frac{1}{2\alpha^{(p)}}\|{\bf u} - {\bf u}^{(p)}\|^2, \label{eq:x_update}\\
{\bf w}^{(p+1)} &= \arg\min_{{\bf w}} 
\mathcal{L}({\bf u}^{(p+1)},{\bf w}^{(p)},{\bm\nu}^{(p)}) 
+ \langle \nabla_{\bf w} \mathcal{L}({\bf u}^{(p+1)},{\bf w}^{(p)},{\bm\nu}^{(p)}), {\bf w} - {\bf w}^{(p)} \rangle
+ \frac{\rho}{2}\|{\bf w} - {\bf u}^{(p+1)}\|^2, \label{eq:w_update}\\
{\bm\nu}^{(p+1)} &= {\bm\nu}^{(p)} + \rho({\bf u}^{(p+1)} - {\bf w}^{(p+1)}). \label{eq:nu_update}
\end{align}
\end{subequations}
\hrulefill 
\vspace*{-4mm} 
\end{figure*}
Specifically, The ${\bf u}$-update in~\eqref{eq:x_update} is separable and admits the closed-form projection:
\begin{align}
[u^{(p+1)}]_i = \left[u^{(p)} - \alpha^{(p)} \left( \nu^{(p)} + \rho(u^{(p)} - w^{(p)}) \right) \right]_i.
\end{align}
The ${\bf w}$-update in~\eqref{eq:w_update} is an unconstrained quadratic problem with solution
\begin{align}
{\bf w}^{(p+1)} = {\bf u}^{(p+1)} - \rho^{-1}(\nabla \phi({\bf u}^{(p+1)}) - {\bm\nu}^{(p)}).
\end{align}
This ADMM variant retains the convergence guarantees of standard ADMM under mild conditions, while significantly reducing the per-iteration complexity compared to SDR.

\subsection{Secure Pinching Beamforming Optimization}
\label{subsec:P_single}
In this subsection, we optimize the pinching beamforming matrix ${\bf \Psi}({\bf P})$ for the single-group multicast secure transmission. The secure transmit beamformer ${\bf w}$ obtained in the previous subsection is assumed fixed here. Therefore, the secrecy multicast rate is purely determined by the PA placements, which can be formulated as follows:
\begin{align} \label{prob:Ps}
\max_{{\bf P}} \ R^{\mathrm{sec}}({\bf P}), \ \ \ \text{s.t.} \ \eqref{S_x}, \ \eqref{delta_min}.
\end{align}
An exhaustive search over the $M^N$ possible configurations is computationally prohibitive even for moderate values of $N$. To reduce the computational complexity and enhance the practical feasibility of the proposed pinching beamforming design, we adopt an element-wise sequential optimization method, where each $x_{m,n} \in \mathcal{S}_x$ is iteratively updated while keeping the other elements fixed.
\subsubsection{Element-wise Problem Reformulation}
When all PAs except the $(m,n)$th one are fixed, \eqref{prob:Ps} reduces to the scalar problem
\vspace{-10pt}
\begin{align}
\max_{x_{m,n}\in{\mathcal S}_{x}} R^{\sec}\big(x_{m,n}\big),
\label{prob:Pn}
\end{align}
where only the $(m,n)$th PA location is variable.  
To explicitly reveal the dependence on $x_{m,n}$, we decompose the aggregated channel into:
\begin{itemize}
\item a fixed term contributed by all other PAs, independent of $x_{m,n}$, and  
\item a variable term contributed solely by the $(m,n)$th PA.
\end{itemize}
For the $k$th Bob, define the fixed part as
\vspace{-5pt}
\begin{align}
S_k^{mn-} &\triangleq \sum_{\substack{q=1 \\ q\neq n}}^{N} 
\frac{\sqrt{\eta}\exp\!\left[-{\rm j}\!\left(k_0D_k(x_{m,q}) + k_{\rm g}x_{m,q}\right)\right] w_m}
{D_k(x_{m,q})}, \label{eq:S_kn}
\end{align}
\vspace{-5pt}
and the variable part as
\begin{align}
A_k(x_{m,n}) &\triangleq \frac{\sqrt{\eta}\exp\!\left[-{\rm j}\!\left(k_0 D_k(x_{m,n}) + k_{\rm g}x_{m,n}\right)\right]w_m}
{D_k(x_{m,n})}, \label{eq:A_kn}
\end{align}
where $D_k(x_{m,n})\triangleq\|\bm\psi_{m,n}-\hat{\bm\psi}_{k}\|$ is the distance between the $(m,n)$th PA and the $k$th Bob.
With this decomposition, the Bob's SINR can be expressed as follows:
\begin{align}
{\rm SINR}_k(x_{m,n}) 
= \frac{P_{\rm t}}{\sigma_{k}^2}\big|S_k^{mn-} + A_k(x_{m,n})\big|^2.
\label{eq:SNRk_single}
\end{align}
An analogous decomposition applies to each Eve, where ${\rm SINR}_\ell(x_{m,n})$ with $S_\ell^{mn-}$ and $A_\ell(x_{m,n})$ defined similarly.

\subsubsection{Equivalent Optimization Objective}
Since $\log(\cdot)$ is monotonic, maximizing $R^{\sec}(x_{m,n})$ is equivalent to maximizing the SINR difference:
\begin{align}
\max_{x_{m,n}\in\mathcal{S}_{x}}
\Big[\min_{k\in\mathcal{K}}{\rm SINR}_k(x_{m,n})  
- \max_{\ell\in\mathcal{L}}{\rm SINR}_{\ell}(x_{m,n}) \Big]^{+}.
\label{eq:pn_obj}
\end{align}

Expanding \eqref{eq:SNRk_single} shows how the fixed and variable parts interact:
\begin{align}
\big|S_k^{mn-} + A_k(x_{m,n})\big|^2 
&= |S_k^{mn-}|^2 + |A_k(x_{m,n})|^2 \notag \\
&+ 2\Re\left\{ S_k^{mn-\,\ast} A_k(x_{m,n}) \right\}, \label{eq:expandBob}
\end{align}
where $|A_k(p_{m,n})|^2 = {\eta}{D_k^{-2}(x_{m,n})}$. Here, the first term $|S_k^{mn-}|^2$ is constant with respect to $x_{m,n}$ and can be ignored in the optimization, which leaves only the second and third terms to be evaluated.
A similar expansion holds for the Eve term. Substituting into \eqref{eq:pn_obj}, the element-wise sequential optimization becomes
\begin{align}\label{eq:expanded_obj}
\max_{x_{m,n}\in\mathcal{S}_{x}}& \Big[\min_{k\in\mathcal{K}}\big(|A_k(x_{m,n})|^2 + 2\Re\{ S_k^{mn-\,\ast} A_k(x_{m,n}) \}\big) \notag \\
&- \max_{\ell\in\mathcal{L}}\big(|A_\ell(x_{m,n})|^2 + 2\Re\{ S_\ell^{mn-\,\ast} A_\ell(x_{m,n}) \}\big) \Big]^{+}.
\end{align}
This decomposition is the key to complexity reduction:  
the contributions of all fixed PAs are pre-computed once, and only the scalar-dependent terms $A_k(\cdot)$ and $A_\ell(\cdot)$ are evaluated in the search for $x_{m,n}$.

\subsubsection{Grid-based Search}
Since \eqref{eq:expanded_obj} involves only a scalar variable over a bounded interval, a uniform grid search provides an efficient solution. We quantize $\mathcal{S}_{x}$ into an $Q$-point grid:
\vspace{-10pt}\begin{align}
\mathcal{S}_{x}^{\rm grid}=\Big\{0,\ \frac{D_{\rm x}}{Q-1},\ \frac{2D_{\rm x}}{Q-1},\ldots,D_{\rm x}\Big\}.
\label{eq:grid_set}
\end{align}
\vspace{-5pt}The optimal position is
\begin{align}
x_{m,n}^{\star} = \arg\max_{x_{m,n}\in\mathcal{S}_{x}^{\rm grid}} \; \widetilde{R}^{\sec}(x_{m,n}),
\label{eq:star}
\end{align}
where $\widetilde{R}^{\sec}(x_{m,n})$ denotes the bracketed term in \eqref{eq:expanded_obj}.
\begin{algorithm}[t]
\caption{AO Algorithm for Solving~\eqref{eq:TxBF_init}}
\label{alg:single}
\begin{algorithmic}[1]
\STATE Initialize ${\bf w}^{(0)}, \mathbf P^{(0)}, \epsilon,\, j \gets 0$
\REPEAT
    \STATE Compute ${\bf w}^{(j+1)}$ by solving~\eqref{eq:standardSDP2} or~\eqref{eq:phi_admm}
    \STATE For each $(m,n)$, update $x_{m,n}$ via~\eqref{eq:expanded_obj}-\eqref{eq:star}
    \STATE $j \gets j + 1$
\UNTIL $\|{\bf P}^{(j+1)} - {\bf P}^{(j)}\| \le \epsilon$ and $\|{\bf w}^{(j+1)} - {\bf w}^{(j)}\| \le \epsilon$
\end{algorithmic}
\end{algorithm}
\vspace{-10pt}\subsection{Complexity and Convergence Analysis}
The proposed AO algorithm for solving~\eqref{eq:Problem-secure} in the single-group scenario is summarized in Algorithm~\ref{alg:single}.
\subsubsection{\bf Complexity Analysis}
For pinching beamforming, the element-wise sequential update incurs a complexity of $\mathcal{O}(MNKLQ)$. 
For transmit beamforming, the SDR-based method has a complexity of $\mathcal{O}(M^6)$, whereas the Dinkelbach-ADMM approach reduces it to $\mathcal{O}(I_{\rm iter} M^3)$, where $I_{\rm iter}$ denotes the number of ADMM iterations. Numerical results in Section~\ref{simulation} confirm that the Dinkelbach-ADMM solution achieves near-SDR performance with significantly lower computational cost.

\subsubsection{\bf Convergence Analysis}
In each AO iteration, the transmit beamformer update via \eqref{eq:standardSDP2} or \eqref{eq:phi_admm} and the pinching beamforming update via \eqref{eq:expanded_obj}-\eqref{eq:star} are both designed to maximize the secrecy multicast rate with the other variable fixed. Therefore, the objective value of problem~\eqref{eq:TxBF_init} is non-decreasing over iterations. Since the secrecy rate is upper-bounded by the finite transmit power and the channel gain constraints, Algorithm~\ref{alg:single} is guaranteed to converge to a stationary solution. 

\section{Joint Secure Beamforming for Multi-Group Multicast Communications}\label{Multiple_group}
In this section, we investigate the joint design of secure transmit beamforming and pinching beamforming for the PASS-based multi-group multicast system. 

\subsection{Problem Reformulation}
We apply the MM surrogate to tackle the non-convexity of $R_{g,k}^{\sec}$ for each subproblem. Specifically, we first define ${\bf W}_k = {\bf w}_k{\bf w}_k^{\rm H}$, then the SINR in~\eqref{sinr_bob} and~\eqref{sinr_eve} can be rewritten as follows:
\begin{align}
{\rm SINR}_{g,k}({\bf W}, {\bf P})
&= \frac{{\bf h}_{g,k}^{\rm T}({\bf P}){\bf W}_g{\bf h}_{g,k}({\bf P})}
{\sum\limits_{j\neq g}{\bf h}_{g,k}^{\rm T}({\bf P}){\bf W}_j{\bf h}_{g,k}({\bf P})+\sigma_{g,k}^2}, \label{eq:SINR_Bob}\\
{\rm SINR}_{g,l}({\bf W},{\bf P})
&= \frac{{\bf h}_{l}^{\rm T}({\bf P}){\bf W}_k{\bf h}_{l}({\bf P})}
{\sum\limits_{j\neq l}{\bf g}_{j}^{\rm T}({\bf P}){\bf W}_j{\bf g}_{j}({\bf P})+\sigma_{{\rm e},j}^2}, \label{eq:SINR_Eve}
\end{align}
The original problem~\eqref{eq:Problem-secure} can be reformulated as follows:
\begin{subequations}\label{prob:P1}
\begin{align}
\max_{\{{\bf W}_g\},{\bf P}} \ 
& R^{\text{sec}}(\{{\bf W}_g\},{\bf P}) \label{P1a}\\
\text{s.t.}\quad 
& \sum_{g=1}^{G}\operatorname{Tr}({\bf W}_g) \le P_{\rm t}, \label{P1b}\\
& {\bf W}_g \succeq {\bf 0},\ \operatorname{rank}({\bf W}_g)=1,\ \forall g\in\mathcal{G}, \label{P1c}\\
& \eqref{S_x}, \ \eqref{delta_min}, \label{P1d}
\end{align}
\end{subequations}
Problem~\eqref{prob:P1} is challenging due to:  
(i) the difference-of-concave (DoC) secrecy rate structure;  
(ii) the rank-one constraints in~\eqref{P1c};  
(iii) the strong coupling between $\{\mathbf{W}_g\}$ and $\mathbf{P}$.
We first apply SDR by dropping the rank-one constraints in \eqref{P1c}. To handle the DoC form, we rewrite the secrecy term in \eqref{P1a} as a sum of concave functions minus concave functions.

Define the following functions 
\begin{align}
F_{1,g,k}({\bf W}, {\bf P}) &= \log_2\!\Big(\operatorname{Tr}\big({\bf H}_{g,k}({\bf P}){\bf W}\big)+\sigma_{g,k}^2\Big),\label{eq:f1}\\
J_{1,g,k}({\bf W}, {\bf P}) &= \log_2\!\Big(\operatorname{Tr}\big({\bf H}_{g,k}({\bf P}){\bf W}^{g-}\big)+\sigma_{g,k}^2\Big),\label{eq:g1}\\
F_{2,g,l}({\bf W}, {\bf P}) &= \log_2\!\Big(\operatorname{Tr}\big({\bf H}_{{\rm e},l}({\bf P}){\bf W}^{g-}\big)+\sigma_{{\rm e},l}^2\Big),\label{eq:f2}\\
J_{2,g,l}({\bf W}, {\bf P}) &= \log_2\!\Big(\operatorname{Tr}\big({\bf H}_{{\rm e},l}({\bf P}){\bf W}\big)+\sigma_{{\rm e},l}^2\Big),\label{eq:g2}
\end{align}
where ${\bf W} \triangleq \sum_{g=1}^{G}{\bf W}_g$, ${\bf W}^{g-}\triangleq \sum_{j\neq g}{\bf W}_g$, ${\bf H}_{g,k}({\bf P})\triangleq {\bf h}_{g,k}({\bf P}){\bf h}_{g,k}^{\rm T}({\bf P})$, and ${\bf H}_{{\rm e},j}({\bf P})\triangleq {\bf h}_{{\rm e}, j}({\bf P}){\bf h}_{{\rm e},j}^{\rm T}({\bf P})$. Then the secrecy multicast rate of the $g$th group admits the following DoC representation:
\begin{align}
R_g^{\sec}({\bf W}, {\bf P}) \ge \min_{k\in\mathcal{K}_g} \big[F_{1,g,k}({\bf W}, {\bf P}) - J_{1,g,k}({\bf W}, {\bf P})\big] \notag \\
 - \max_{l\in\mathcal{L}}\big[F_{2,g,l}({\bf W}, {\bf P}) - J_{2,g,l}({\bf W}, {\bf P})\big]. \label{eq:SR_DoC}
\end{align}

To handle the non-convex DoC structure in~\eqref{eq:SR_DoC}, we adopt the MM framework within an AO strategy. Specifically, each concave term appearing with a negative sign, such as $J_{1,g,k}$ and $J_{2,g,l}$, is replaced by its first-order affine upper bound at the current iterate. Let $\widetilde{{\bf W}}$ and $\widetilde{\bf P}$ denote the current feasible points for ${\bf W}$ and ${\bf P}$, respectively. The generic linearization takes the form as follows:
\begin{align}
J({\bf Z}) \le J(\widetilde{{\bf Z}}) + \mathrm{Tr}\!\left(\nabla J(\widetilde{{\bf Z}})^{{\rm H}}({\bf Z} - \widetilde{{\bf Z}})\right) 
\triangleq \widehat{J}({\bf Z} \,|\, \widetilde{{\bf Z}}), \label{eq:MM_generic}
\end{align}
where ${\bf Z}$ represents either ${\bf W}$ or ${\bf P}$.
Substituting~\eqref{eq:MM_generic} into the secrecy multicast rate constraints yields a convex surrogate problem for the current AO framework. This MM-based AO scheme is then applied alternately to update ${\bf W}$ and ${\bf P}$ until convergence.

\subsection{Secure Transmit Beamforming Design}\label{subsec:W_update}
\subsubsection{MM-Based SDR Formulation}
In the ${\bf W}$-update step, the pinching beamforming ${\bf P}$ is fixed. Applying the MM surrogate~\eqref{eq:MM_generic} to $J_{1,g,k}({\bf W})$ and $J_{2,g,l}({\bf W})$ yields
\begin{figure*}[!t]
\normalsize
\begin{align}
J_{1,g,k}({\bf W})
&\le J_{1,g,k}(\widetilde{\bf W}) + \sum_{i=1}^{G} \operatorname{Tr}\Big(\nabla_{{\bf W}_i} J_{1,g,k}(\widetilde{{\bf W}}) ({\bf W}_i-\widetilde{\bf W}_i)\Big)
\triangleq \widehat{J}_{1,g,k}({\bf W}|\widetilde{\bf W}), \label{eq:j1_hat}\\
J_{2,g,l}({\bf W})
&\le J_{2,g,l}(\widetilde{\bf W}) + \sum_{i=1}^{G} \operatorname{Tr}\Big(\nabla_{{\bf W}_i} J_{2,g,l}(\widetilde{\bf W}) ({\bf W}_i-\widetilde{\bf W}_i)\Big)
\triangleq \widehat{J}_{2,g,l}({\bf W}|\widetilde{\bf W}). \label{eq:j2_hat}
\end{align}
\hrulefill 
\vspace*{-4mm} 
\end{figure*}
where $\nabla_{{\bf W}_i}(\cdot)$ denotes the gradient with respect to ${\bf W}_i$.

Replacing $J_{1,g,k}$ and $J_{2,g,l}$ in~\eqref{eq:g1} and~\eqref{eq:g2} with their surrogates~\eqref{eq:j1_hat} and~\eqref{eq:j2_hat} gives the convex objective:
\begin{subequations}\label{prob:P2W}
\begin{align}
& \max_{\{{\bf W}_i \succeq \mathbf{0}\}_{i=1}^G, \, t} \quad t \label{P2W_a}\\
&\mathrm{s.t.} \quad
F_{1,g,k}({\bf W}) + F_{2,g,l}({\bf W}) \notag\\
&\qquad - \widehat{J}_{1,g,k}({\bf W} \,|\, \widetilde{{\bf W}}) - \widehat{J}_{2,g,l}({\bf W} \,|\, \widetilde{{\bf W}}) \ge t, \label{P2W_b}\\
&\qquad\forall k\in\mathcal{K}_g,\ \forall g\in\mathcal{G},\ \forall l\in \mathcal{L}, \notag\\
&\qquad\eqref{P1b}.
\end{align}
\end{subequations}
Problem~\eqref{prob:P2W} is a standard semidefinite program and can be efficiently solved using CVX. The resulting $\{{\bf W}_g\}$ is then used in the subsequent pinching beamformer update step of the AO framework.

\subsubsection{Low-Complexity SOCP-Based Algorithm}\label{subsec:SOCP}
To reduce the computational complexity of solving~\eqref{prob:P2W}, we develop a SOCP-based algorithm. Let $\xi_{{\rm b}(g,k)}$ and $\xi_{{\rm e}(g,l)}$ denote the lower and upper bounds of ${\rm SINR}_{g,k}$ and ${\rm SINR}_{g,l}$, respectively, i.e.,
\begin{align}
{\rm SINR}_{g,k} \ge \xi_{{\rm b}(g,k)},  \quad 
{\rm SINR}_{g,l} \le \xi_{{\rm e}(g,l)}.
\end{align}
Introducing a secrecy slack variable $t$, problem~\eqref{prob:P2W} can be equivalently reformulated as follows:
\begin{subequations}
\label{prob:P3}
\begin{align}
& \max_{\{{\bf w}_g\},\,{\bf P},\,t,\,\boldsymbol{\xi}_{\rm b},\,\boldsymbol{\xi}_{\rm e}} \quad
{\rm min} \ t \label{P3a}\\
\text{s.t.}\quad 
& {\rm SINR}_{g,k} \ge \xi_{{\rm b}(g,k)}, \quad \forall g\in\mathcal{G},\ \forall k\in\mathcal{K}_g, \label{P3b}\\
& {\rm SINR}_{g,l} \le \xi_{{\rm e}(g,l)}, \quad \forall g\in\mathcal{G}, \forall l\in\mathcal{L}, \label{P3c}\\
& 1+\xi_{{\rm b}(g,k)} \ge 2^{t}\big(1+\xi_{{\rm e}(g,l)}\big), \quad \forall g,\ \forall k,\ \forall l, \label{P3d}\\
& \eqref{P1b}, \ \eqref{S_x}, \ \eqref{delta_min}. \label{P3e}
\end{align}
\end{subequations}
Constraints~\eqref{P3b} and~\eqref{P3c} can be expressed as fractional quadratic inequalities. Specifically, \eqref{P3b} is equivalent to
\begin{equation}
\sum_{j\neq g}\big|{\bf h}_{g,k}^{\rm T} {\bf w}_j\big|^2 + \sigma_{g,k}^2 
\ \le\ \frac{\big|{\bf h}_{g,k}^{\rm T} {\bf w}_g\big|^2}{\xi_{{\rm b}(g,k)}},\quad \forall g,k,
\label{eq:rearrange_b}
\end{equation}
and \eqref{P3c} is equivalent to
\begin{equation}
\big|{\bf h}_{l}^{\rm T}{\bf w}_g\big|^2 
\ \le\ \xi_{{\rm e}(g,k)} \Big(\sum_{j\neq g}\big|{\bf h}_{l}^{\rm T}{\bf w}_j\big|^2 + \sigma_{{\rm e},l}^2\Big),\quad \forall g,l.
\label{eq:rearrange_e}
\end{equation}

It is well known that for any complex scalar $x$ and positive scalar $r$, the following first-order inequality holds:
\begin{equation}
\frac{|x|^2}{r} \ \ge\ 
\frac{2\,{\rm Re}\{\tilde{x}^\ast x\}}{\tilde{r}} 
- \frac{|\tilde{x}|^2}{\tilde{r}^2}\, r 
\ \triangleq\ B(x, \ r; \ \tilde{x}, \ \tilde{r}), 
\label{eq:F_linear}
\end{equation}
where $(\tilde{x},\tilde{r})$ is a fixed point. Following the inequality, the right-hand side of \eqref{eq:rearrange_b} can be lower-bounded by $B\big({\bf h}_{g,k}^{\rm T} {\bf w}_g,\; \xi_{{\rm b}(g,k)};\; {\bf h}_{g,k}^{\rm T} \tilde{{\bf w}}_g,\; \tilde{\xi}_{{\rm b}(g,k)}\big)$. Applying~\eqref{eq:F_linear} to~\eqref{eq:rearrange_b} yields the convex approximation
\vspace{-5pt}
\begin{equation}
\sum_{j\neq g}\big|{\bf h}_{g,k}^{\rm T} {\bf w}_j\big|^2 + \sigma_{g,k}^2\le 
B\Big({\bf h}_{g,k}^{\rm T}{\bf w}_g, \xi_{{\rm b}(g,k)}; {\bf h}_{g,k}^{\rm T}\tilde{{\bf w}}_g,\tilde{\xi}_{{\rm b}(g,k)}\Big).
\label{eq:approx_b}
\end{equation}
Similarly, treating $\xi_{{\rm e}(g,l)}$ as fixed within one SOCP iteration, \eqref{eq:rearrange_e} can be conservatively approximated as follows:
\begin{align}
\frac{\big|{\bf h}_{l}^{\rm T}{\bf w}_g\big|^2}{\xi_{{\rm e}(g,k)}} & \le \sum_{j\neq g} B\Big({\bf h}_{l}^{\rm T}{\bf w}_j,\ 1;\ {\bf h}_{l}^{\rm T}\tilde{{\bf w}}_j,\ 1\Big) + \sigma_{{\rm e},l}^2.\label{eq:approx_e}
\end{align}

With \eqref{eq:approx_b} and \eqref{eq:approx_e}, the non-convex constraints \eqref{P3b} and \eqref{P3c} are replaced by convex inequalities. Together with \eqref{P3d} and the power budget \eqref{P3e}, we obtain the following SOCP
\begin{subequations}
\label{prob:P3-1prime}
\begin{align}
\max_{\{{\bf w}_g\},\,\boldsymbol{\xi}_{\rm b},\,\boldsymbol{\xi}_{\rm e},\,t} 
& {\rm min} \ \ t \label{P3-1prime_a}\\
\text{s.t.} \ \ & \eqref{eq:approx_b},\ \eqref{eq:approx_e},\ \eqref{P3d},\ \eqref{P3e}. \label{P3-1prime_b}
\end{align}
\end{subequations}
Problem~\eqref{prob:P3-1prime} can be efficiently solved via CVX. In the outer AO loop, $\{\boldsymbol{\xi}_{\rm b},\boldsymbol{\xi}_{\rm e}\}$ are updated based on the current $\{\mathbf{w}_g\}$ and reused in the next SOCP iteration until convergence.

\subsection{Pinching Beamforming Optimization}\label{subsec:P_multi}
In this subsection, we address the optimization of the pinching beamforming matrix $\mathbf{\Psi}({\bf P})$ under the MM-based AO framework. When $\{{\bf W}_g\}$ is fixed, the secrecy-rate constraints in~\eqref{prob:P1} remain non-convex with respect to ${\bf P}$ due to the concave terms $J_{1,g,k}({\bf P})$ and $J_{2,g,l}({\bf P})$ appearing with negative signs. Following the MM principle, we replace these concave terms with their first-order upper bounds evaluated at the current feasible point $\widetilde{{\bf P}}$.

Specifically, for each $(g,k)$ and $(g,l)$, the following inequalities hold:
\begin{align}
J_{1,g,k}({\bf P}) 
&\le J_{1,g,k}(\widetilde{\bf P}) 
+ \frac{ {\rm Tr}\left[ \left( {\bf H}_{g,k}({\bf P}) - \widetilde{\bf H}_{g,k} \right) {\bf W}^{g-} \right] }
{{\ln 2}\left( {\rm Tr}\left( \widetilde{\bf H}_{g,k} {\bf W}^{g-} \right) + \sigma_{g,k}^2 \right)} \notag \\
&\triangleq \widehat{J}_{1,g,k}({\bf P} \,|\, \widetilde{\bf P}), \label{eq:J1_upper_correct}
\end{align}
and 
\begin{align}
J_{2,g,l}({\bf P}) 
&\le J_{2,g,l}(\widetilde{\bf P}) 
+ \frac{ {\rm Tr}\left[ \left( {\bf H}_{{\rm e},l}({\bf P}) - \widetilde{\bf H}_{{\rm e},l} \right) {\bf W} \right] }
{{\ln 2}\left( {\rm Tr}\left( \widetilde{\bf H}_{{\rm e},l} {\bf W} \right) + \sigma_{{\rm e},l}^2 \right)} \notag \\
&\triangleq \widehat{J}_{2,g,l}({\bf P} \,|\, \widetilde{\bf P}), \label{eq:J2_upper_correct}
\end{align}
where $\widetilde{\bf H}_{g,k} \triangleq {\bf H}_{g,k}(\widetilde{\bf P})$ and $\widetilde{\bf H}_{{\rm e},l} \triangleq {\bf H}_{{\rm e},l}(\widetilde{\bf P})$.

Substituting \eqref{eq:J1_upper_correct} and \eqref{eq:J2_upper_correct} into the secrecy multicast rate constraints, the MM surrogate problem for the transmit beamformer update can be written as follows:
\begin{subequations}
\label{prob:P_Pupdate}
\begin{align}
\max_{{\bf P},\,t} \quad & t \label{P_Pupdate_a}\\
\mathrm{s.t.} \quad 
& F_{1,g,k}({\bf P}) + F_{2,g,l}({\bf P}) \notag\\
& - \widehat{J}_{1,g,k}({\bf P}|\widetilde{\bf P}) - \widehat{J}_{2,g,l}({\bf P}|\widetilde{\bf P}) \ge t, \ \forall g,k,l, \label{P_Pupdate_b}\\
& \eqref{S_x},\ \eqref{delta_min}. \notag
\end{align}
\end{subequations}
Problem~\eqref{prob:P_Pupdate} is still non-convex due to the non-linear dependence of ${\bf H}_{g,k}({\bf P})$ on each variable $x_{m,n}$. To enable low-complexity optimization, we adopt an element-wise sequential update strategy: at each inner iteration, all entries of $\mathbf{P}$ are fixed except one $x_{m,n}$, and \eqref{prob:P_Pupdate} is solved over $x_{m,n}$ only.

For a specific $x_{m,n}$, the key quadratic term can be expressed as follows:
\begin{align}
{\rm Tr}\left( {\bf H}_{g,k}(x_{m,n}){\bf W}_g \right) 
= \left| S_{g,k}^{mn-} + A_{g,k}(x_{m,n}) \right|^2, \label{eq:quad_expand}
\end{align}
where
\begin{align}
& S_{g,k}^{mn-}\triangleq \sum_{m = 1}^{M}\sum_{\substack{q=1\\q\neq n}}^{N}\frac{\sqrt{\eta}\exp\left[{-{\rm j}}\big(k_0D_{g,k}(x_{m,q}) + k_{\rm g}x_{m,q}\big)\right][{\bf w}_g]_m}{D_{g,k}(x_{m,q})}, \\
& A_{g,k}(x_{m,n})\triangleq \frac{\sqrt{\eta}\exp\big[-{\rm j}\left(k_0 D_{g,k}(x_{m,n}) + k_{\rm g}x_{m,n}\right)\big][{\bf w}_g]_m}{D_{g,k}(x_{m,n})}.
\end{align}
Similar expansions apply to the eavesdropper terms. 

Subsequently, the MM-based element-wise sequential optimization for pinching beamforming can be reformulated as follows:
\begin{align}\label{eq:multiple-PA}
	\max_{x_{m,n}}\min  \ \widetilde{R}^{\sec}(x_{m,n}), \ {\rm s.t.}  \  \eqref{S_x}, \ \eqref{delta_min}
\end{align}
where
\vspace{-10pt}
\begin{align}
	\widetilde{R}^{\sec}(x_{m,n})& = F_{1,g,k}(x_{m,n}) + F_{2,g,l}(x_{m,n})\notag \\
	& - \widehat{J}_{2,g,l}(x_{m,n}|\widetilde{\bf P}) - \widehat{J}_{2,g,l}(x_{m,n}|\widetilde{\bf P}).
\end{align}
\vspace{-5pt}
and
\begin{align}
	& F_{1,g,k}(x_{m,n})  = \log_2\Big(\sum_{g = 1}^G\big|S_{g,k}^{mn-} + A_{g,k}(x_{m,n})\big|^2 + \sigma_{g,k}^2\Big) \\
	& \widehat{J}_{1,g,k}(x_{m,n}|\widetilde{\bf P})  = {\rm const}_{g,k} + \frac{ \sum_{j \neq g}^G\big|S_{j,k}^{mn-} + A_{j,k}(x_{m,n})\big|^2}{{\ln 2} \operatorname{Tr}\Big( {\bf H}_{g,k}(\widetilde{\bf P}) {\bf W}^{g-} \Big) + \sigma_{g,k}^2} \\
	& {\rm const}_{g,k}  = J_{1,g,k}(\widetilde{\bf P}) - \frac{ \operatorname{Tr}\Big({\bf H}_{g,k}(\widetilde{\bf P}){\bf V}^{g-} \Big)}{{\ln 2} \operatorname{Tr}\Big( {\bf H}_{g,k}(\widetilde{\bf P}) {\bf W}^{g-} \Big) + \sigma_{g,k}^2}.
\end{align}
\vspace{-5pt}

The Eve terms $F_{2,g,l}(x_{m,n})$, $\widehat{J}_{2,g,l}(x_{m,n}|\widetilde{\bf P})$ and ${\rm const}_{g,l}$ can be expressed as the similar forms. Since $\mathcal{S}_x$ is a finite feasible set determined by~\eqref{S_x} and~\eqref{delta_min}, the optimal $x_{m,n}$ can be efficiently found. By calculating and fixing the constant terms $\left\{S_{g,k}^{mn-}, S_{g,l}^{mn-}, {\rm const}_{g,k}, \ {\rm const}_{g,l}\right\}$, $\forall g,k,l$ beforehand, the optimal $x_{m,n}$ can be effectively obtained via a low-complexity one-dimensional grid-based search over the limited feasible set ${\mathcal S}_x^{\rm grid}$. The solution $x_{m,n}$ is then adopted for the next iteration. 
\begin{algorithm}[t]
\caption{AO Algorithm for Solving~\eqref{prob:P1}}
\label{alg:multiple}
\begin{algorithmic}[1]
\STATE initialize ${\bf W}^{(0)}, {\bf P}^{(0)}, \epsilon,\, j \gets 0$

\REPEAT
    \STATE compute ${\bf W}^{(j+1)}$ by solving~\eqref{prob:P2W} or~\eqref{prob:P3-1prime}
    \STATE update ${\bf P}^{(j+1)}$ by solving~\eqref{eq:multiple-PA}
    \STATE $j \gets j + 1$
\UNTIL $\|{\bf P}^{(j+1)} - {\bf P}^{(j)}\| \le \epsilon$ and $\|{\bf W}^{(j+1)} -{\bf W}^{(j)}\| \le \epsilon$
\end{algorithmic}
\end{algorithm}
\vspace{-10pt}
\subsection{Complexity and Convergence Analysis}
The proposed AO algorithm for solving~\eqref{eq:Problem-secure} in the multi-group scenario is summarized in Algorithm~\ref{alg:multiple}.
\subsubsection{\bf Complexity Analysis}
For transmit beamforming, the MM-based SDR method incurs a computational complexity of $\mathcal{O}(G^6 M^6)$, where $G$ denotes the number of multicast groups. The SOCP algorithm is developed as a low-complexity alternative, with complexity $\mathcal{O}((GM)^{3.5} \log(1/\epsilon))$, where $\epsilon$ is the target accuracy. The pinching beamformer is optimized via MM-based sequential element-wise updates, yielding a complexity of $\mathcal{O}(MNGKLQ)$.

\subsubsection{\bf Convergence Analysis}
The surrogate function~\eqref{prob:P3} generates a monotonically non-decreasing sequence of objective values for problem~\eqref{prob:P1}. Since the optimal value of~\eqref{prob:P1} is upper-bounded by the transmit power and the channel gain constraints, Algorithm~\ref{alg:multiple} is guaranteed to converge to a stationary solution.
\vspace{-5pt}
\section{Numerical Results}\label{simulation}
We now validate the effectiveness of the proposed PASS framework for secure multicast communications through comprehensive numerical simulations. Specifically, we compare the performance of PASS with conventional fixed-location multi-antenna baselines under the same simulation setup. 

\subsection{Simulation Setup}
For the configuration of PASS, the waveguides are deployed at a height of $h = 5$ m with a total length of $D_{\rm x}$. The number of discrete search points in the one-dimensional grid based search is set to $Q = 10^3$. The PAs are deployed along the waveguide with a minimum separation of $\Delta_{\rm min} = \lambda/2$. The effective refractive index is assumed to be $n_{\rm eff} = 1.44$ \cite{10945421}.

For the multicast configuration, the carrier frequency is set to $f_{\rm c} = 28$ GHz, and the noise power at each Bob and Eve is assumed to be $\sigma^2 = -90$ dBm. In the multi-group case, the group number is set as $G= 2$. Unless otherwise specified, the users are assumed to be uniformly distributed within the service region. User grouping is assumed to be random throughout the optimization. The design of user grouping strategies is beyond the scope of this article and is left for future work. All numerical results are obtained by averaging over $1000$ independent random channel realizations.

For the ADMM algorithm, the penalty parameter $\rho$ and the step size $\alpha^{(p)}$ are initialized as $\rho = 4L_{\phi}$ and $\alpha^{(p)} = \frac{8}{37L_{\phi}}, \ \forall p$, where $L_{\phi}$ is the Lipschitz constant of $\nabla\phi({\bf w})$~\cite{li2019constant}. For the LSE-based algorithm, the smoothing parameter is set to $\beta=10$. The maximum iteration numbers of the Dinkelbach method and ADMM are both set to $I_{\rm iter} = 50$. For the outer AO framework alternating between the pinching beamforming and transmit beamforming designs, the maximum iteration number is set to $50$. The convergence tolerances are $\epsilon = 10^{-3}$. The PA location matrix ${\bf P}$ is initialized by randomly selecting $MN$ discrete positions that satisfy the MC constraint. The initialized transmit beamforming matrix ${\bf W}$ has entries independently drawn from a circularly symmetric complex Gaussian distribution, and is normalized to satisfy the total transmit power constraint.
\subsection{Baseline Architectures}
We compare the proposed PASS architecture with two representative MIMO baselines: \textit{massive MIMO} and \textit{conventional MIMO}. The configurations are summarized as follows:  
\begin{itemize}    
    \item \textbf{\textit{Massive MIMO}}: A fully digital architecture with $MN$ antennas, each driven by an individual RF chain. The beamformer is designed using the Dinkelbach-based SDR and MM-based SDR algorithms for the single-group and multi-group scenarios, respectively. While this architecture can achieve high DoFs, its substantial number of RF chains leads to significantly higher hardware complexity and energy consumption compared with both the conventional MIMO and the proposed PASS architectures.  
    
    \item \textbf{\textit{Conventional MIMO}}: A traditional fully digital system equipped with $M$ antennas. Each antenna is connected to a dedicated RF chain. The beamformer is designed using the same algorithms as in the massive MIMO case (i.e., Dinkelbach-based SDR and MM-based SDR), which ensures a fair comparison in terms of precoding strategy.
\end{itemize}

For all baselines, the antenna elements are arranged in a half-wavelength spaced uniform linear array centered at $[D_{\rm x}/2, \ 0, \ h]^{\rm T}$ and aligned along the $y$-axis.

\begin{figure}[!t]
\centering
\includegraphics[height=0.26\textwidth]{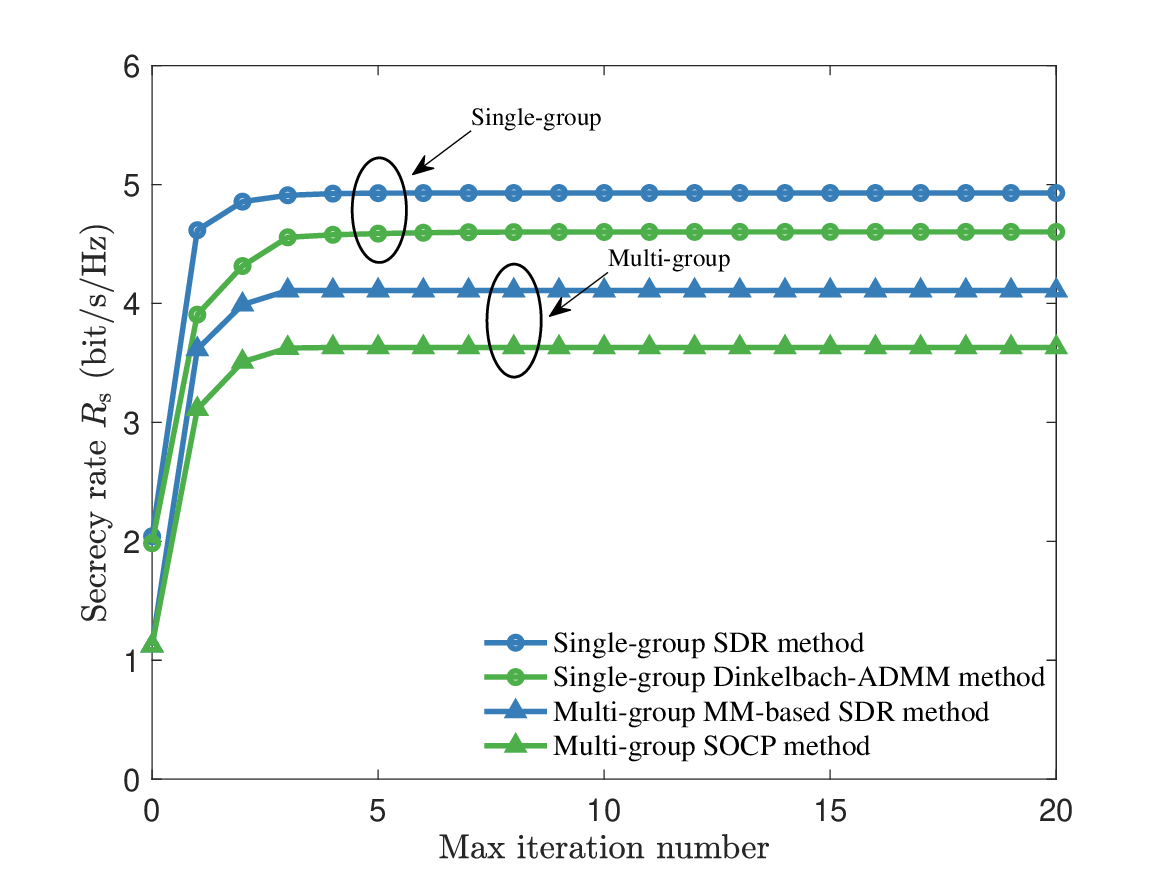}
\caption{Multicast rate versus the number of max iteration in AO algorithms, where $D_{\rm x} = 20$ m, $D_{\rm y} = 6$ m, $(M,N) = (8, 4)$, $(K, L) = (2,2)$ and $P_{\rm t} = -20$dBm.}
\label{convergence}
\vspace{-10pt}
\end{figure}
\subsection{Convergence of the AO-Based Joint Beamforming}
{\figurename}\ref{convergence} depicts the convergence behavior of Algorithms~\ref{alg:single} and Algorithms~\ref{alg:multiple}, where the transmit and pinching beamformers are alternately optimized. In both single- and multi-group multicast scenarios, the achievable secrecy rate increases rapidly and stabilizes within a small number of iterations, which confirms the fast convergence of the proposed AO framework.
Although both algorithms exhibit similar convergence speed, the achievable secrecy rate in the multi-group scenario is lower than that in the single-group case. This performance degradation arises from the inter-group interference and the fairness-driven objective in the multi-group formulation, which maximizes the minimum secrecy rate among all groups and thus inherently limits the overall system throughput.

In the single-group case, the SDR-based transmit beamforming achieves the highest performance but incurs prohibitive complexity due to large-scale semidefinite programming, whereas the Dinkelbach-ADMM method offers a significantly more scalable alternative with minor performance loss. For multi-group multicast, the MM-based SDR approach reveals favorable performance but scales poorly with the number of users and groups, while the SOCP-based method avoids matrix lifting and achieves a favorable tradeoff between complexity and performance. Pinching beamformers in both scenarios are optimized in an sequential element-wise method, which ensures monotonic convergence with substantially reduced complexity compared to exhaustive search and facilitates efficient joint optimization in large-scale PASS-enabled multicast secrecy networks.

\subsection{Single-Group Multicast Scenario}
\subsubsection{Secrecy multicast rate versus the transmit power}
\begin{figure}[!t]
\centering
\includegraphics[height=0.26\textwidth]{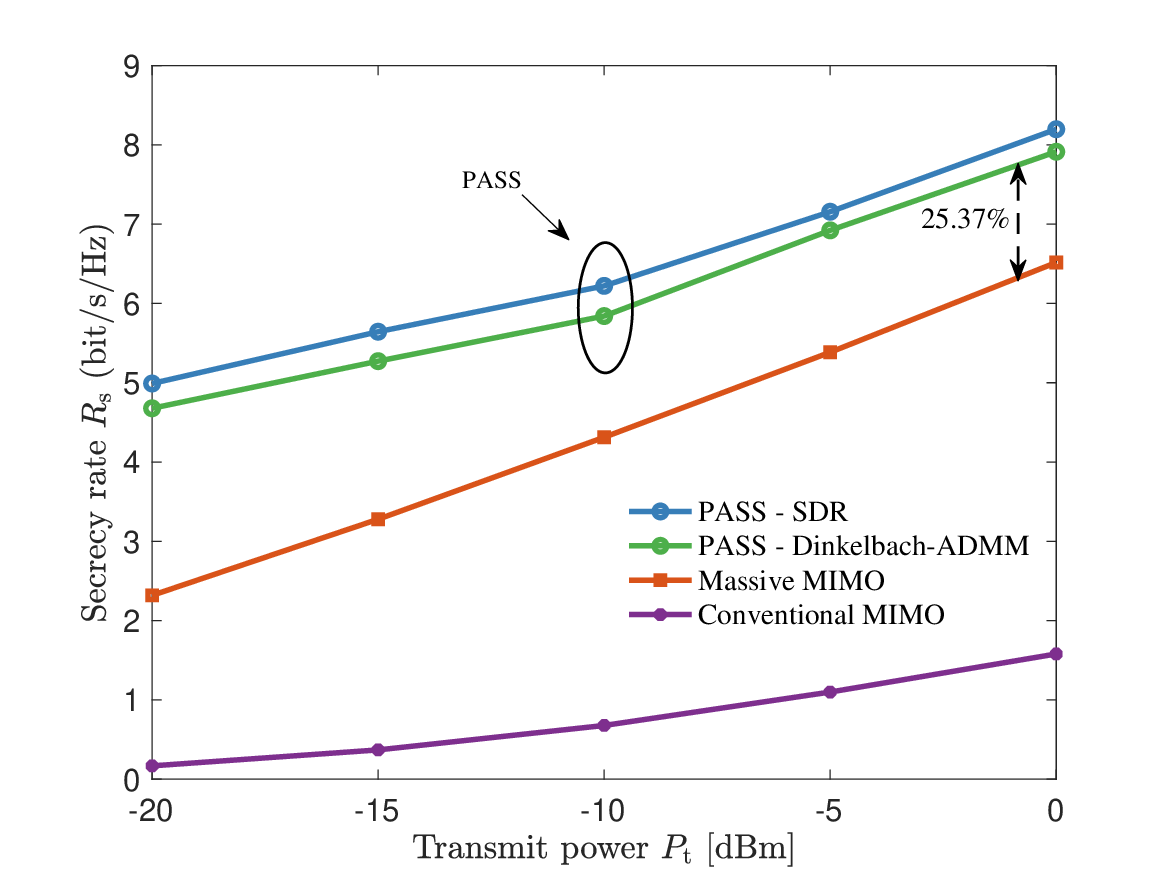}
\caption{Single-group secrecy multicast rate versus the transmit power with $D_{\rm x} = 20$ m, $D_{\rm y} = 6$ m, $(M, N) = (8, 4)$, and $(K, L) = (4, 4)$.}\label{sg:power}
\vspace{-5pt}
\end{figure}
{\figurename}~{\ref{sg:power}} illustrates the secrecy multicast rate performance of the proposed single-group schemes as a function of the transmit power $P_{\rm t}$. As expected, the secrecy rate increases monotonically with the transmit power across all considered schemes. Among the proposed methods, the SDR-based PASS algorithm achieves the best performance, as it provides an upper-bound solution by relaxing the non-convex constraints. Notably, the ADMM-based algorithm offers comparable performance to SDR while substantially reducing computational complexity. This observation confirms the effectiveness of the ADMM-based method in approximating the performance of high-complexity algorithms while maintaining scalability.

Moreover, we observe that the performance gain brought by the proposed PASS architecture becomes less pronounced as the transmit power increases. This is because, the system performance becomes increasingly constrained by the precoder's ability to distinguish Bobs from Eves, rather than by signal strength alone. While PASS benefit from flexible antenna positioning, massive MIMO inherently provides more DoFs, which helps suppress interference and leakage more effectively in high-SNR regions. Nonetheless, the proposed PASS architecture still significantly outperforms all conventional fixed-location antenna systems, which validates its practical advantages.

\subsubsection{Secrecy multicast rate versus the deployment region}
\begin{figure}[!t]
\centering
\includegraphics[height=0.26\textwidth]{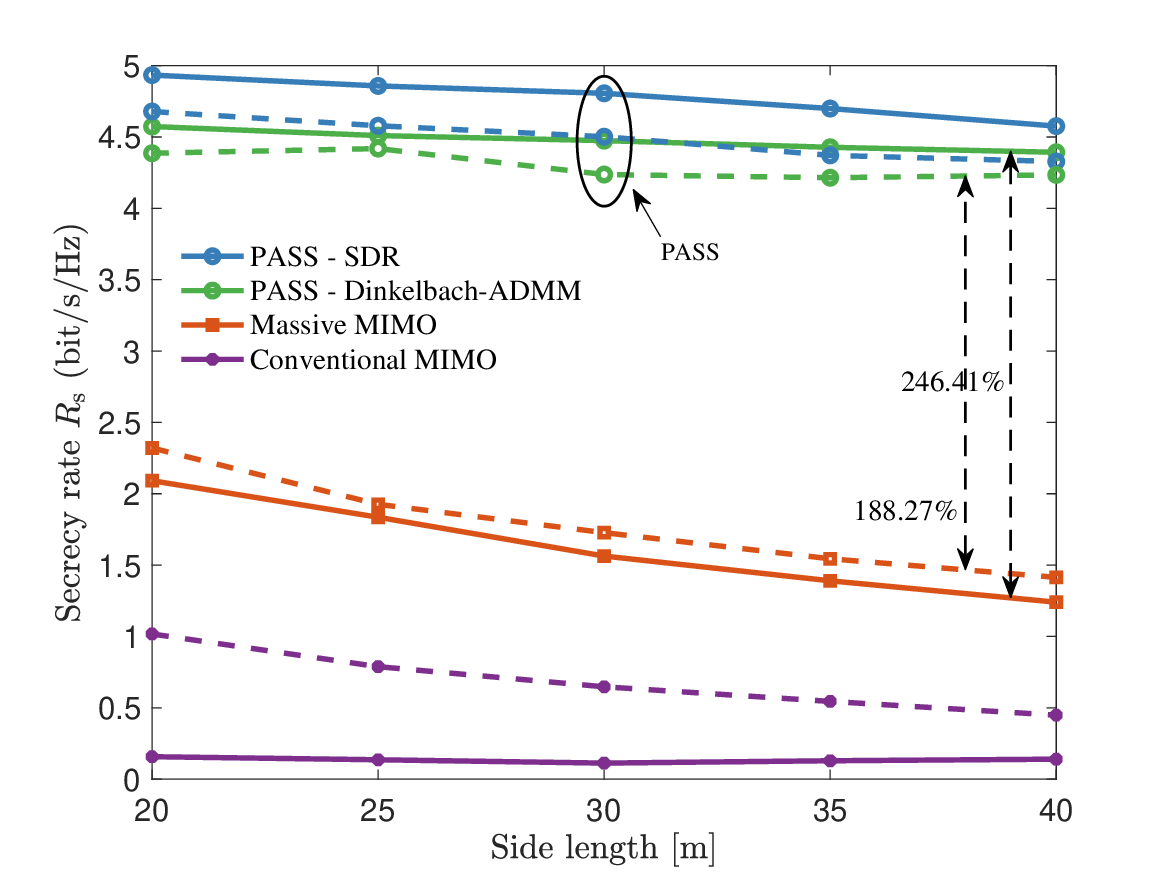}
\caption{Single-group secrecy multicast rate versus the side length $D_{\rm x}$ under two service regions: $D_{\rm y} = 6$ m (solid lines) and $D_{\rm y} = 30$ m (dashed lines). $P{\rm t} = -20$ dBm, $(M, N) = (8, 4)$, and $(K, L) = (4, 4)$.}\label{sg:region}
\vspace{-10pt}
\end{figure}
{\figurename}~{\ref{sg:region}} illustrates the secrecy multicast rate under different spatial deployment configurations. Obviously, the proposed PASS architecture significantly outperforms all baseline fixed-location antenna systems. Although the secrecy multicast rate of PASS decreases slightly as the side length $D_{\rm x}$ increases, its degradation is considerably slower compared to fixed-location systems. This robustness is due to the fact that multicast performance is inherently constrained by the worst-case Bobs. In conventional antenna deployments, expanding the coverage area leads to longer path lengths and increased path loss. In contrast, PASS adaptively adjusts PA positions to ensure stable LoS distances, which mitigates performance degradation even under large deployment sizes. Moreover, a comparison between the $D_{\rm y} = 6$~m and $D_{\rm y} = 30$~m scenarios reveals that the performance gain of PASS over traditional architectures becomes more pronounced as the deployment scale grows. This suggests that the benefits of PASS in larger service region.

\subsubsection{Secrecy multicast rate versus the PASS deployment}
\begin{figure}[!t]
\centering
\includegraphics[height=0.26\textwidth]{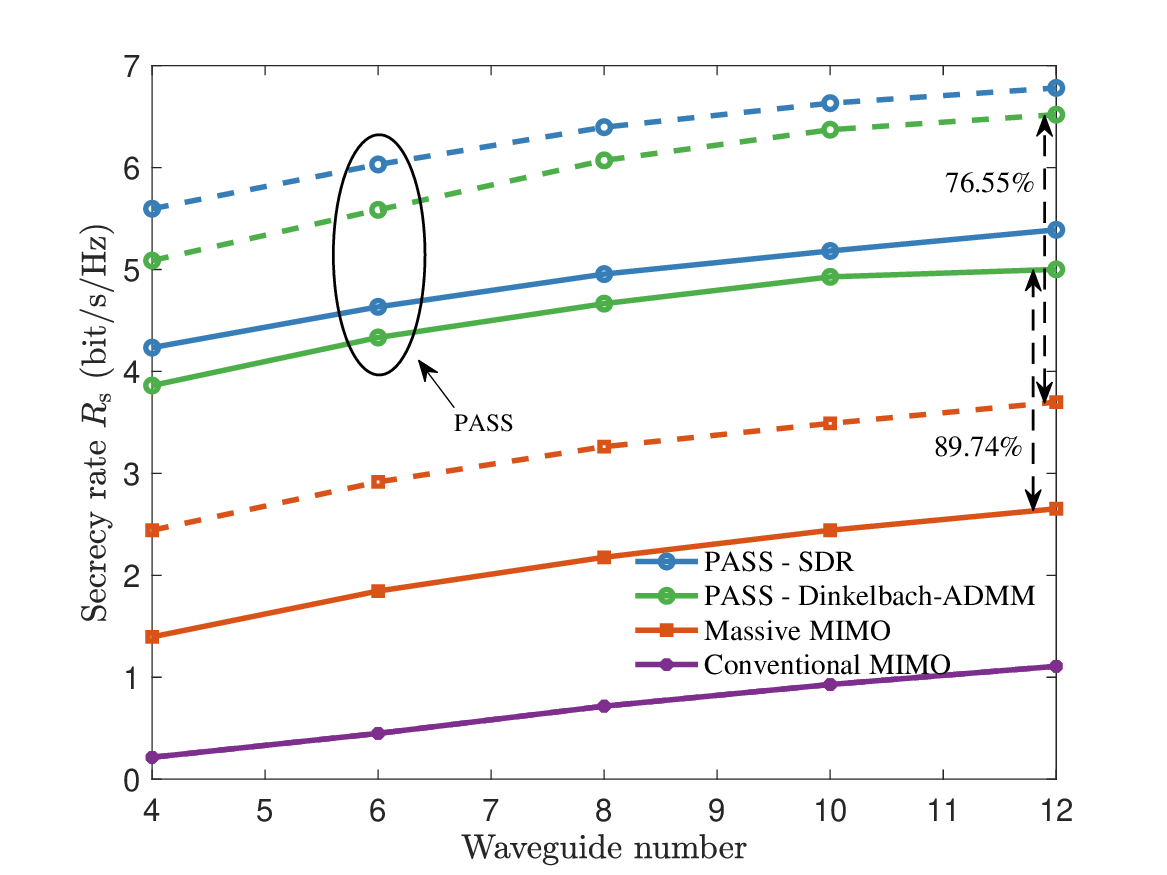}
\caption{Single-group secrecy multicast rate versus the number of waveguide under two PASS configuration: $N = 4$ (solid lines) and $N = 10$ (dashed lines). $P{\rm t} = -20$ dBm, $D_{\rm x} = 20$ m, $D_{\rm y} = 6$ m, and $(K, L) = (4, 4)$.}\label{sg:passnumber}
\vspace{-15pt}
\end{figure}
{\figurename}~{\ref{sg:passnumber}} illustrates the secrecy multicast rate as a function of the number of waveguides under two different PA configurations. As shown, the secrecy rate of all schemes increases steadily with the number of waveguides, which reflects the benefit of increased spatial resolution and aperture size in the pinching beamforming. Another observation is that the performance gap between ADMM and SDR remains nearly constant across different waveguide numbers. This implies that the ADMM-based algorithm retains its near-optimal performance even as the array size scales up. Given its substantially lower computational complexity, this robustness further validates ADMM as an effective and practical alternative to high-complexity SDR in large-scale deployments.

In addition, while the overall PASS gain over fixed-location antenna systems is stable with respect to waveguide number, the performance improvement becomes more pronounced under a larger PA configuration. This highlights the scalability of PASS in antenna-limited systems, where the joint effect of additional PAs and waveguides enables stronger spatial adaptability. The observed gain further demonstrates that PASS are especially beneficial in massive MIMO scenarios, where large antenna arrays demand both reconfigurability and beamforming precision.

\subsubsection{Secrecy multicast rate versus the user density}
\begin{figure}[!t]
\centering
\includegraphics[height=0.26\textwidth]{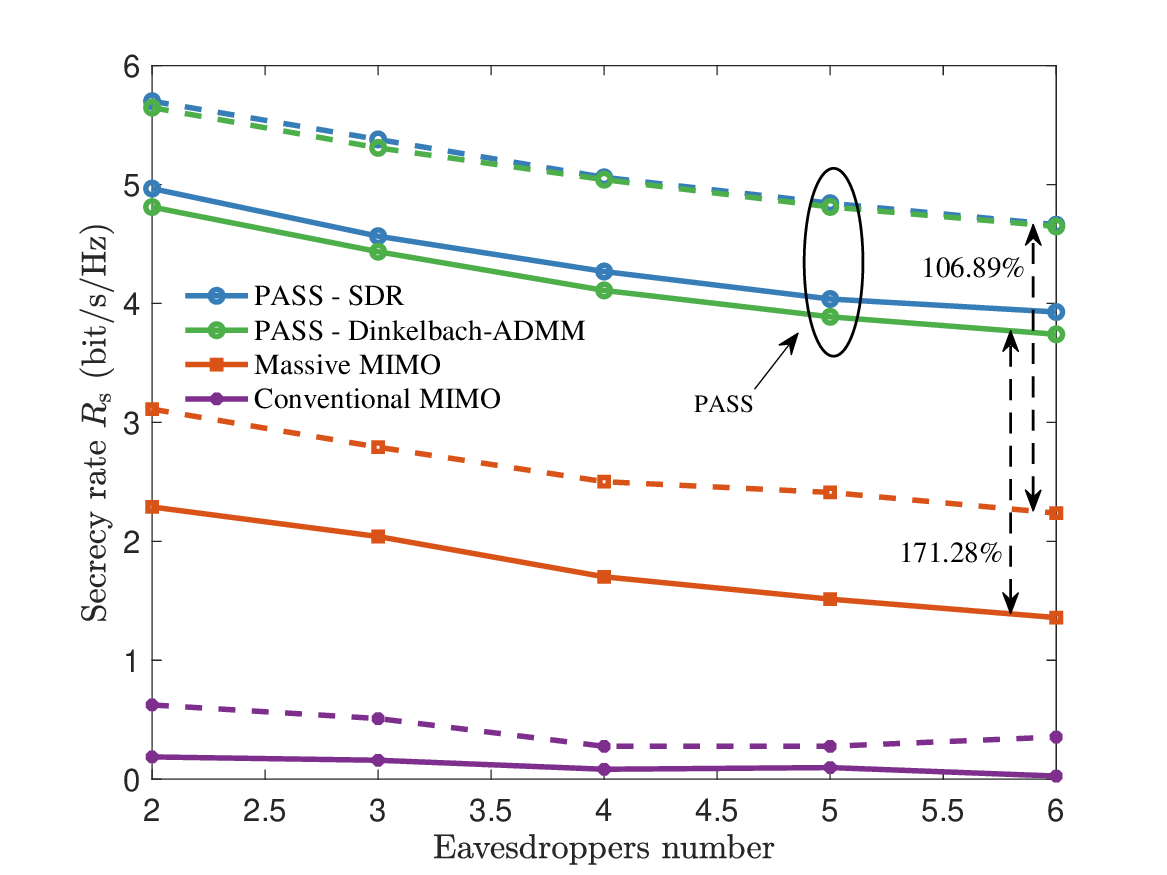}
\caption{Single-group secrecy multicast rate versus the number of Eve under two Bob numbers: $K = 4$ (solid lines) and $K = 2$ (dashed lines). $P{\rm t} = -20$ dBm, $D_{\rm x} = 20$ m, and $(M, N) = (8, 4)$.}\label{sg:usernumber}
\vspace{-5pt}
\end{figure}
{\figurename}~{\ref{sg:usernumber}} demonstrates the secrecy rate performance as a function of the number of eavesdroppers. Several performance characteristics can be observed from the figure. First, it is evident that the proposed PASS schemes consistently outperform all baseline fixed-location antenna architectures across all scenarios, which indicates strong robustness against increasing security threats. As the number of eavesdroppers increases, the secrecy rate of all schemes declines due to enhanced information leakage. Nevertheless, the PASS architecture exhibits a significantly slower degradation trend, which reflects its superior ability to spatially separate the Bobs and Eves via LoS channel reconfiguration. Second, the performance gain of PASS over conventional fixed-location schemes becomes more significant with larger $K$, which reinforces that PASS are particularly well-suited for high-density multicast scenarios, where spatial user separation and reconfigurability are critical for preserving secrecy guarantees. Third, while the SDR-based method provides an upper-bound performance, the low-complexity Dinkelbach-ADMM solution achieves nearly identical secrecy rates, which further validates its effectiveness. 

\subsection{Multi-group Multicast Scenario}
\subsubsection{Secrecy multicast rate versus the transmit power}
\begin{figure}[!t]
\centering
\includegraphics[height=0.26\textwidth]{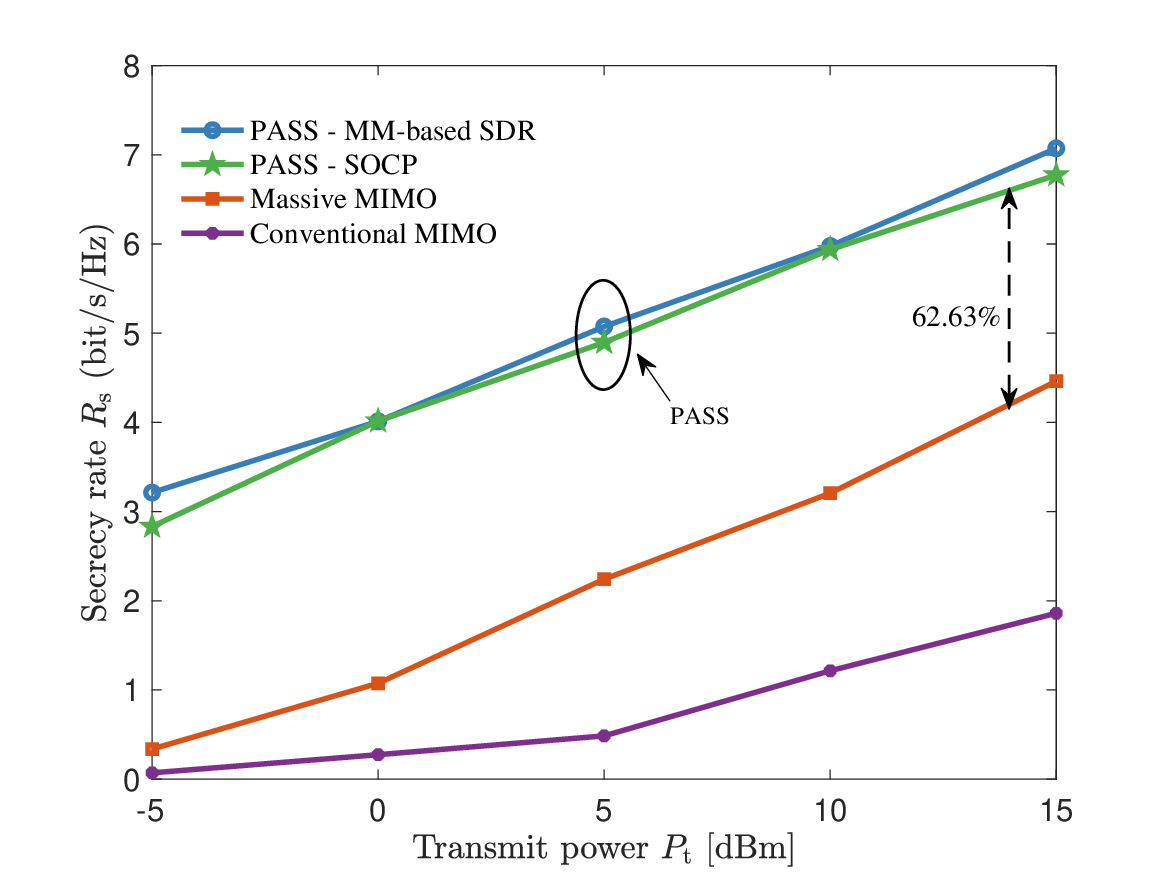}
\caption{Multi-group secrecy multicast rate versus the transmit power with $D_{\rm x} = 20$ m, $D_{\rm y} = 6$ m, $(M, N) = (8, 4)$, and $(K, L) = (4, 4)$.}\label{mg:power}
\vspace{-15pt}
\end{figure}
{\figurename}~{\ref{mg:power}} illustrates the secrecy rate performance of various schemes in the multi-group multicast setting as a function of the transmit power $P_{\rm t}$. It can be observed that the proposed PASS architecture significantly outperforms all fixed-location antenna benchmarks over the entire SNR range. Moreover, the PASS schemes using SOCP and MM-based SDR algorithms achieve nearly identical secrecy rate performance. This close performance highlights the effectiveness of the SOCP-based formulation in approximating the upper-bound solution while offering substantially reduced computational complexity. 

Furthermore, although increasing transmit power benefits all schemes, the performance gap between PASS and conventional fixed-location systems remains consistently large. This implies that the spatial adaptability of PASS provides persistent advantages even in high-SNR regimes. While massive MIMO systems provide higher spatial degrees of freedom, the performance gains of PASS stem from its dynamic reconfiguration capability. By repositioning the PAs in response to user locations, PASS facilitate efficient spatial separation between Bobs and Eves. These results reinforce the suitability of PASS for secrecy in multicast systems.

\subsubsection{Secrecy multicast rate versus the deployment region}
\begin{figure}[!t]
\centering
\includegraphics[height=0.26\textwidth]{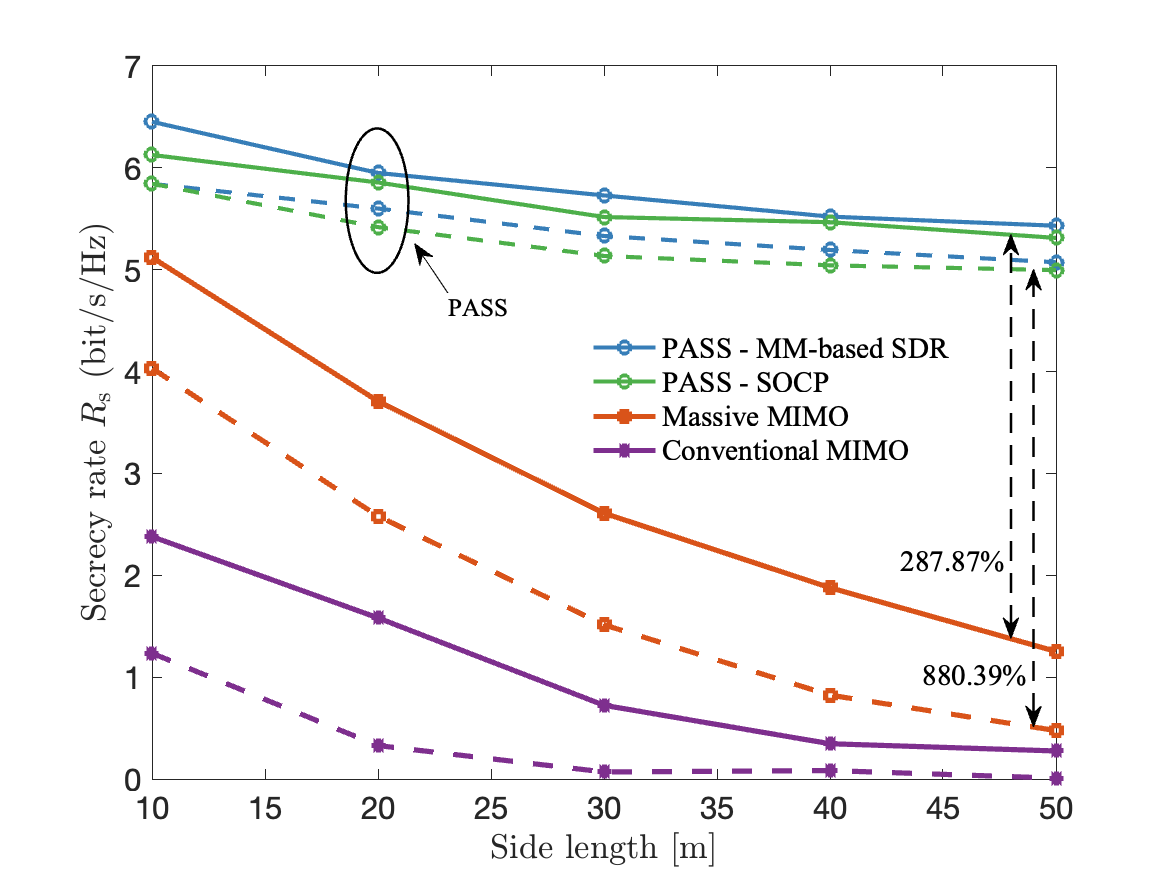}
\caption{Multi-group secrecy multicast rate versus the side length $D_{\rm x}$ under two service regions: $D_{\rm y} = 6$ m (solid lines) and $D_{\rm y} = 20$ m (dashed lines). $P{\rm t} = 0$ dBm, $(M, N) = (8, 4)$, and $(K, L) = (4, 4)$}\label{mg:region}
\vspace{-5pt}
\end{figure}
{\figurename}~{\ref{mg:region}} presents the secrecy rate performance of different schemes in a multi-group multicast scenario as a function of the deployment region $D_{\rm x}$. While the SDR-based approach provides an upper-bound performance benchmark, the proposed low-complexity SOCP solution achieves near-identical secrecy rates with substantially reduced computational burden.

Consistent with the observations in the single-group case, the PASS architecture demonstrates remarkable robustness against increased coverage size. Specifically, as $D_{\rm x}$ grows, the performance of PASS degrades only slightly. This stems from its capability to dynamically reposition PAs that effectively reduces the average distance between transmit elements and Bobs, which mitigates the large-scale fading. 
Additionally, we compare system performance under different deployment dimensions $D_{\rm y}$. The results reveal that the performance advantage of PASS becomes even more pronounced with a larger $D_{\rm y}$, which further emphasizes the spatial scalability of the proposed architecture. These observations confirm that PASS are particularly well-suited for large-scale secrecy multicast deployments, where both coverage adaptability and secrecy performance are essential.

\subsubsection{Secrecy multicast rate versus the PASS deployment}
\begin{figure}[!t]
\centering
\includegraphics[height=0.26\textwidth]{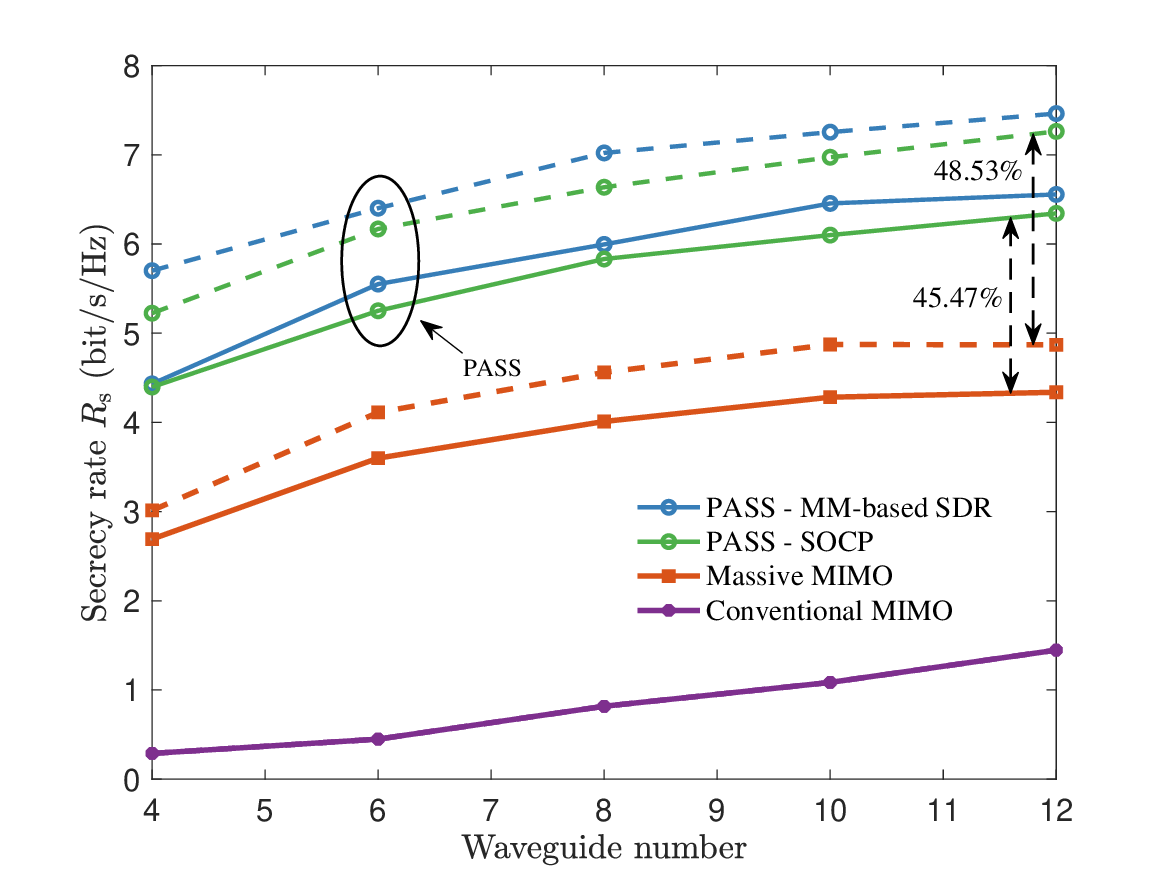}
\caption{Multi-group secrecy multicast rate versus the number of waveguide under two PASS configuration: $N = 4$ (solid lines) and $N = 10$ (dashed lines). $P{\rm t} = 0$ dBm, $D_{\rm x} = 20$ m, $D_{\rm y} = 6$ m, and $(K, L) = (4, 4)$.}\label{mg:passnumber}
\vspace{-15pt}
\end{figure}
{\figurename}~{\ref{mg:passnumber}} illustrates the secrecy rate performance of various antenna systems in a multiple-group multicast scenario as a function of the number of waveguides. As the number of waveguides increases, the MM-SDR based method provides performance upper bounds but incurs higher computational complexity. In comparison, the SOCP approach offers a lower-complexity alternative with scalable implementation, while maintaining competitive secrecy performance.

It is observed that increasing the number of waveguides leads to a consistent improvement in system performance. This is attributed to the enhanced spatial DoFs provided by additional waveguides, which improves the precoding capability of the PASS architecture and facilitates more effective separation between Bobs and Eves. Moreover, the performance gap between PASS and conventional fixed-location antenna systems becomes more pronounced as the number of PAs increases. 
Moreover, even under identical waveguide deployments, the gain achieved by increasing PAs in PASS surpasses the performance improvements observed in massive MIMO through the addition of DoFs. These results suggest that PASS not only maintains low hardware complexity and power consumption, but also achieves superior secrecy performance through adaptive antenna placement and optimized pinching beamforming, especially in dense multi-group deployments.

\subsubsection{Secrecy multicast rate versus the user density}
\begin{figure}[!t]
\centering
\includegraphics[height=0.26\textwidth]{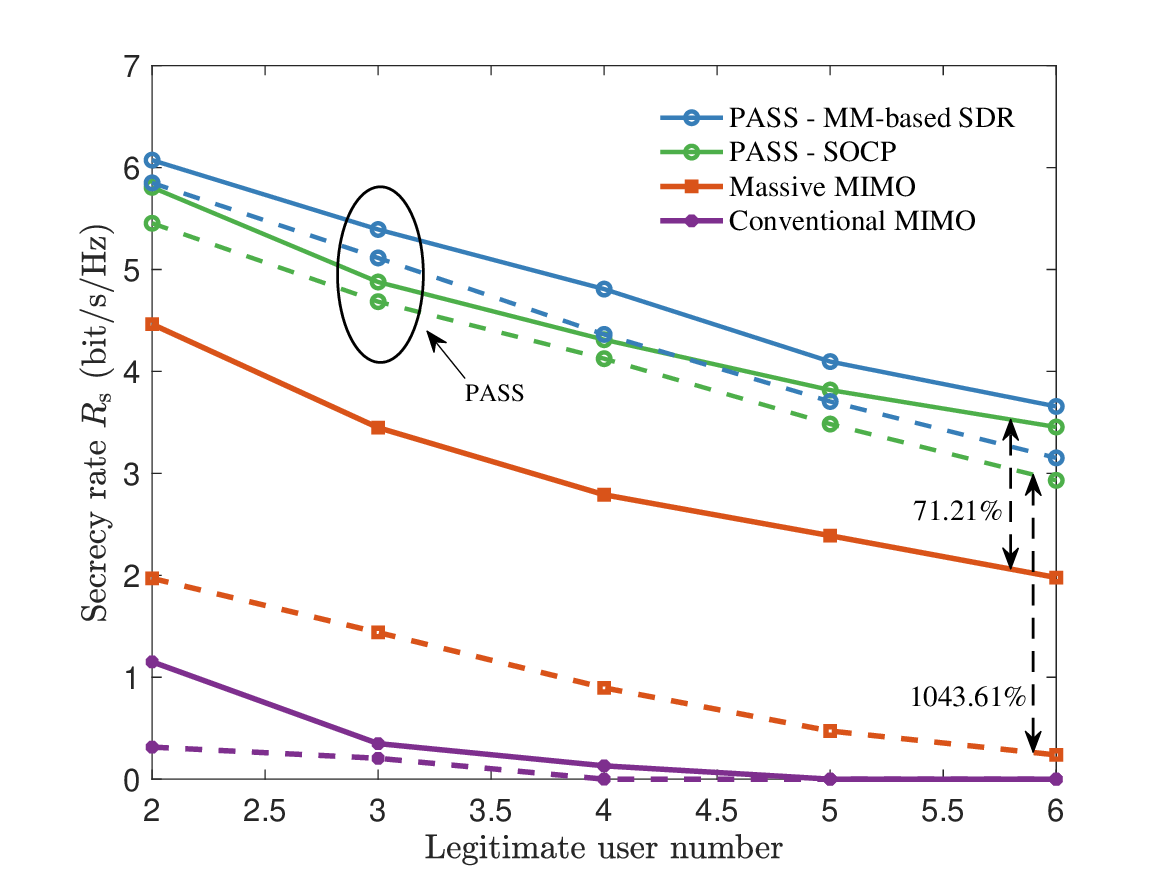}
\caption{Multi-group secrecy multicast rate versus the number of Bob under two Eve numbers: $L = 4$ (solid lines) and $L = 2$ (dashed lines). $P{\rm t} = 0$ dBm, $D_{\rm x} = 20$ m, and $(M, N) = (8, 4)$.}\label{mg:usernumber}
\vspace{-5pt}
\end{figure}
{\figurename}~{\ref{mg:usernumber}} illustrates the impact of the number of users on the secrecy performance in a multiple-group multicast scenario. While the MM-SDR solution serves as a performance reference, the SOCP-based method demonstrates better scalability with respect to user density.
As the number of users increases, a noticeable degradation in secrecy rate is observed across all schemes due to increased inter-group interference. Despite this, the proposed PASS architecture consistently outperforms all baseline fixed-location antenna systems.
Furthermore, we compare system robustness under different levels of Eve threats. It is observed that the performance of PASS degrades relatively slowly with an increasing number of Eves, whereas conventional massive MIMO schemes suffer significantly. This performance resilience stems from the spatial reconfigurability of PASS, which dynamically adjusts the positions of PAs to shorten the distance to intended users and enhance beamforming isolation between Bobs and Eves. These findings underscore the scalability and adaptability of the PASS architecture in high user-density scenarios.

\section{Conclusions}\label{conclusion}
 This paper presented a unified secrecy transmission framework for multicast communications in PASS, which jointly optimizes transmit and pinching beamforming to maximize the secrecy multicast rate. The framework accommodates both single-group and multi-group scenarios with multiple legitimate users and a common set of eavesdroppers.
For the single-group case, we developed an AO framework, where the pinching beamformer is updated via an element-wise sequential optimization. The transmit beamformer is solved either through SDR after Charnes-Cooper transformation or via a low-complexity alternative based on the Dinkelbach method with ADMM-based subproblem updates. For the multi-group case, we adopted an MM framework, under which the pinching beamformer is similarly updated, while the transmit beamformer is obtained via SDR or an efficient SOCP formulation.
Simulation results confirm that PASS significantly outperforms conventional fixed-location antenna systems in terms of secrecy performance. The performance gains become more pronounced with increasing coverage size, antenna array size, and user density. These results demonstrate that PASS serve as a cost-effective and energy-efficient solution for scalable multicast secrecy in large-scale or dense user scenarios.
\bibliographystyle{IEEEtran} 
\bibliography{reference}    

\end{document}